\documentclass{iopart}
\usepackage{graphicx}

\expandafter\let\csname equation*\endcsname\relax
\expandafter\let\csname endequation*\endcsname\relax

\usepackage{amsmath}
\usepackage{amsfonts}
\usepackage[dvips]{epsfig}
\usepackage{bm}

\usepackage[T1]{fontenc}
\usepackage[latin9]{inputenc}
\usepackage{float}
\usepackage{amsmath}
\usepackage{graphicx}
\usepackage{amssymb}


\begin{document}

\title{Quantum Simulations of Lattice Gauge Theories using Ultracold Atoms in Optical Lattices}

\author{Erez Zohar}
\address{Max-Planck-Institut f\"ur Quantenoptik, Hans-Kopfermann-Stra\ss e 1, 85748 Garching, Germany.}

\author{J. Ignacio Cirac}
\address{Max-Planck-Institut f\"ur Quantenoptik, Hans-Kopfermann-Stra\ss e 1, 85748 Garching, Germany.}

\author{Benni Reznik}
\address{School of Physics and Astronomy, Raymond and Beverly Sackler Faculty of Exact Sciences, Tel Aviv University, Tel-Aviv 69978, Israel.}

\begin{abstract}
Can high energy physics  be simulated by  low-energy, non-relativistic, many-body systems, such as ultracold atoms?
Such ultracold atomic systems lack the type of symmetries and dynamical properties of high energy physics models: in particular, they manifest neither local gauge invariance nor Lorentz invariance, which are crucial properties of the quantum field theories which are the building blocks of the standard model of elementary particles.

However, it turns out, surprisingly, that there are ways to configure atomic system to manifest both local gauge invariance and Lorentz invariance. In particular, local gauge invariance can arise either as an effective, low energy, symmetry, or as an "exact" symmetry, following from the conservation laws in atomic interactions.  Hence, one could hope that such quantum simulators may lead to new type of (table-top) experiments, that shall be used to study various QCD phenomena, as the confinement of dynamical quarks, phase transitions, and other effects, which are inaccessible using the currently known computational methods.

In this report, we review the Hamiltonian formulation of lattice gauge
theories, and then describe our recent progress in constructing quantum simulation of Abelian and non-Abelian lattice gauge theories
in $1+1$ and $2+1$ dimensions using ultracold atoms in optical lattices.

\end{abstract}

\maketitle

\tableofcontents{}

\pagebreak{}

\section{Introduction: Quantum Simulation of High Energy Physics}

\subsection{Quantum Simulations}

Quantum simulation \cite{CiracZoller} is a relatively new physical
field, concerning the simulation of quantum systems with other quantum
systems. The original idea goes back to Richard Feynman \cite{Feynman1982},
who argued that since nature is quantum, it should be simulated by
quantum computers. However, it was "left aside" for about two
decades, as no candidates for playing the role "quantum simulators"
were available. Following Feynman's pioneering work \cite{Feynman1982},
it has been shown that universal quantum computers \cite{Lloyd1996,Zalka1998}
could in principle simulate general many-body systems \cite{Lidar1997}
as well as quantum fields \cite{Jordan2012}, serving as digital quantum
simulators.

Thanks to the enormous progress carried out over the last decades
on the road towards "Universal Quantum Computers", several quantum
systems are now highly and precisely controllable in the laboratory,
mainly belonging to atomic physics and quantum optics. These are,
for example, ultracold atoms in optical lattices \cite{Jaksch1998,Lewenstein2007,Bloch2008,Bloch2012,Lewenstein2012},
including Bose-Einstein condensates; Trapped ions \cite{james,Leibfried2003,Lanyon2011,Blatt2012,Schneider2012};
photonic systems \cite{Aspuru-Guzik2012}, Rydberg Atoms \cite{Rydberg}
and more. Thus, although a full-fledged, universal quantum computer
has not been realized yet, one may utilize these systems in order
to create \emph{analog quantum simulators}, which are indeed non-universal
simulators, but nevertheless, allow for quantum simulation of quantum
systems. This is done by mapping the degrees of freedom of the simulated
systems into those of the controllable simulating system. Such quantum
simulators provide simple realizations for quantum models and phenomena
which may be, otherwise, inaccessible, either theoretically, experimentally,
or numerically, amendable for direct observation of many-body physics
phenomena. Using such simulators, one could study, experimentally,
the spectrum and dynamics of the simulated system, and therefore these
simulators may be regarded as analog quantum computers, specifically
tailored to the simulated systems. Quantum simulation also has the
possibility of realizing physical models which are "unreal"
- not believed to be found in nature.

One could also propose and realize \emph{digital quantum simulators},
which fulfill the same tasks, but using quantum computation methods
of qubits and quantum gates. This work, however, deals with analog
simulations.

Thus, quantum simulation is currently a rapidly growing physical area,
involving multidisciplinary research, both theoretical and experimental.
The theoretical research focuses on mapping the simulated physics
into the quantum simulating systems, considering possible realizations
and required approximations of the simulated model, as well as methods
to control and get the most out of the simulating systems. Hence the
multidisciplinarity: it involves both the simulated and the simulating
physics, which may belong to two totally different physical disciplines,
at least from a traditional perspective. The experimental research
focuses on improving the capabilities and controllability of the simulating
systems, also by developing new techniques and simulating systems,
and of course, on realizing quantum simulations, serving as table-top
experiments,\emph{ }with great success.

Having been realized using AMO
 physics, models of Condensed Matter and statistical mechanics are
the natural candidates for quantum simulations. These may be Hubbard
models, spin chains and Heisenberg models and others - only few examples are \cite{Jaksch1998,Demler2002,Greiner2002,Garcia2003,Stoferle2004,porras2004,Garcia2004,Lewenstein2004,Jaksch2005,Wehr2006,Ortner2009,Britton2012,Endres2012}. Effects of artificial, external gauge potentials,
experienced by cold atoms, have also been considered for quantum simulation,
among more topological effects, such as the Aharonov-Bohm effect,
Berry's phase, topological insulators and the quantum Hall effect
\cite{Jaksch2003,Mueller2004,Jaksch2005,Sorensen2005,Osterloh2005,Ruseckas2005,Juzeliunas2008,Paredes2008,Goldman2009,Boada2010,Alba2011,Aidelsburger2013,Miyake2013}.
A significant example to the power of quantum simulation in condensed
matter physics is the experiment for simulating the phase transition
in Hubbard models \cite{Greiner2002} .

However, quantum simulations are not limited only to condensed matter
physics, and they involve also other physical areas, such as gravity,
including Black-Hole Hawking \cite{Hawking1974} and Unruh \cite{Unruh1976} radiation
\cite{Garay2001,Barcelo2001,Fischer2004,Visser2005,Retzker2008,Horstmann2010,Steinhauer2010,Horstmann2011,Steinhauer2013}.
Relativistic physics has been considered as well, including, for example,
Dirac Equation (including Zitterbewegung and the Klein paradox) \cite{Lamata2005,Vaishnav2008,Gerritsma2010,Gerritsma2011},
the famous Fermi two-atom problem \cite{FP,Sabin2011}, neutrino oscillations
\cite{Noh2012} and the Majorana equation \cite{Casanova2011,Noh2013}.
The next obvious step is to consider quantum field theories as well.

Quantum simulations of quantum field theories involve these of vacuum
entanglement of a scalar field \cite{Reznik2003,Retzker2005} as well as the
interacting scalar-fermionic theories of Thirring and Gross-Neveu
\cite{Cirac2010}. Fermions in lattice gauge theories have been considered
as well, for example in the quantum simulations of Axions and Wilson
fermions \cite{Bermudez2010}, Dirac fields in curved spacetime \cite{Boada2011},
general simulations of QFT with topological insulators \cite{Mazza2012}
and Schwinger's pair-creation mechanism \cite{Szpak2012}. Simulations
of nontrivial geometries and topologies and extra
dimensions, using internal degrees of freedom \cite{Boada2012,Boada2014}
have been suggested too, as well as simulations of some supersymmetric theories,
such as \cite{Yu2008,Shi2010,Yang2010}.

\subsection{Quantum Simulations of High Energy Physics}

Another class of quantum simulations is those of high energy physics
(HEP), which involve dynamical gauge fields.

Gauge fields are in the core of the standard model of particle physics
\cite{Halzen1984,Langacker2011}. These fields, through their local
symmetry, induce the interactions among matter particles, which are
fermions
 - or, phrased differently, the excitations of the gauge fields are
the so called gauge bosons, force carriers, or interaction mediators.
(In the current report, we shall not discuss the Brout-Englert-Higgs mechanism \cite{Brout1964,Higgs1964},
which introduces a scalar field, not playing the role of a gauge field.)

Each gauge theory is based on local gauge invariance - invariance
to local gauge transformations, generated by the so called \emph{gauge
group}. Such groups may be either Abelian or non-Abelian: for example,
$QED$ (Quantum Electrodynamics) is an Abelian gauge theory, based
on the group $U(1)$, while the strong interactions are described
by the theory of $QCD$ (Quantum Chromodynamics), which has the non-Abelian
local gauge symmetry of $SU(3)$. These symmetries induce the interactions
among matter particles, which have different \emph{gauge charges}
- the electric charge in $QED$, or the non-Abelian color charge in
$QCD$, for example. In the case of non-Abelian groups, the gauge
bosons (e.g., $QCD$'s gluons) also carry charge of their own, unlike
the Abelian photons of $QED$.

The standard model of particle physics involves a local gauge symmetry
of $SU\left(3\right)\times SU\left(2\right)\times U\left(1\right)$,
corresponding to the strong ($QCD$), weak and electromagnetic ($QED$)
interactions respectively. All the gauge groups in this case are continuous
groups - Yang-Mills theories \cite{Yang1954}.

The study of gauge theories, within and beyond the framework of the standard model,
employs many theoretical techniques. These include, of course, the
perturbative approach of Feynman diagrams \cite{Bjorken1965,Itzykson1980,Peskin1995,Ramond1997,Srednicki2007},
which has been applied with great success to quantum electrodynamics. However,
within $QCD$, such perturbative methods face a problem. In high energies, or short
distances (such as the scale of deep inelastic scattering) perturbative techniques
work fairly well, since, due to asymptotic freedom \cite{Gross1973}
(and the earlier results of the parton model and Bjorken scaling \cite{Bjorken1969b,Bjorken1969}),
in this regime the quarks behave essentially as free, pointlike particles  (within
the hadrons). On the other hand,
at low energies, or large distances, the quarks are subject to the
so-called \emph{quark confinement} \cite{Wilson}  which forbids the existence of free
quarks and rather binds them together into composite particles - hadrons.
This limit is nonperturbative, and other techniques
must be applied in order to study it.

The most powerful approach that has been developed for non-perturbative $QCD$
effects is
lattice gauge theory \cite{Wilson,KogutLattice,Kogut1983}.
It
allows probing and calculating many important quantities and phenomena
of the theory, both analytically, as the lattice suggests a regularization method,
and numerically, mostly using  Monte-Carlo classical simulation methods.
Such studies have led to significant and remarkable results, such as
the low-energy spectrum of $QCD$ and the CKM matrix elements \cite{FLAG2013},
several results concerning the quark-gluon plasma \cite{Shuryak1980,Svetitsky,McLerran1986,SvetitskyReview}
and the deconfinement phase transition at finite temperature \cite{Polyakov1978,Susskind1979,McLerran1981} - just to name a few.

Monte Carlo calculations, nevertheless, are limited in certain important case.
For example - the computationally
hard sign problem \cite{Troyer2005} that limits calculations in regimes
with a finite chemical potential for the fermions - which is problematic,
for example, for the study of the $QCD$ phases of color-superconductivity
or quark-gluon plasma \cite{McLerran1986,Kogut2004,Fukushima2011},
or the fact that the Monte-Carlo calculations are of Euclidean
correlation functions, and not of real-time dynamics. Indeed a full study
of quark confinement with dynamical charges, and other nonperturbative
phenomena in $QCD$ in 3+1 dimensions is still lacking.

Another successful avenue for the exploration of such theories has been
the development of "toy models" - simpler physical
theories which capture the essential physics in question.
Such studies include, for example,
 confinement in Abelian theories in various dimensions
(including on the lattice) \cite{Schwinger1962I,Schwinger1962,Lowenstein1971,Casher1974,Wilson,KogutLattice,Polyakov,BanksMyersonKogut,DrellQuinnSvetitskyWeinstein,BenMenahem},
 or $QCD_{2}$, the 1+1 dimensional version of
Quantum Chromodynamics, which has been a target of interest over the
last decades, as a natural playground for nonperturbative calculations,
and as a basis for analytical understanding of real-world, 3+1 $QCD$
($QCD_{4}$). There, several nonperturbative methods have been proven
fruitful and useful for derivation of the spectrum of $SU(N_{c})$
theories. First, in the large $N_{c}$ limit, firstly discussed (and
solved) by 't Hooft \cite{tHooft1974,Callan1976}, where the mesonic
spectrum of the theory is revealed, manifesting confinement of quarks.
Other methods to study the hadronic spectrum of $QCD_{2}$ have been
DLCQ (discretized light-cone quantization) \cite{Hornbostel1990} and current algebras \cite{Armoni2001}.
Witten's non-Abelian bosonization \cite{Witten1984} has helped to
gain insight on the strong coupling Baryonic spectrum and quark content
\cite{Date1987,Frishman1987,Frishman1990,Frishman1993}. The relation
of the fermion mass with confinement in these theories has been studied,
for example, in \cite{Gross1996,Armoni1998}.

Thus, new methods of calculations in gauge theories, in particular
$QCD$, could be of a great help. Quantum simulation methods may be candidates
for that. Once a quantum simulator for a gauge theory is built, one
could use it to observe, experimentally, the otherwise inaccessible
physics of these special, nonperturbative regimes. Such quantum simulators
may be the quantum computers of HEP calculations: allowing for real-time dynamics
and including real fermions rather than Grassman variables, suffering from the sign-problem,
they suggest a way to overcome the problems of numerical lattice gauge theories mentioned above. Besides that, reconstructing
the standard model (in this case, out of atomic building blocks),
or any of its sub-models, could help in understanding the basic interactions
and symmetries of the theory.

Quantum simulation of high energy physics, involving gauge field(s),
may be both theoretically and experimentally harder than simulating other physical phenomena, in condensed matter physics, for example.
In fact, three basic requirements \cite{AngMom} must be
met in order to obtain a quantum simulator of a gauge theory:
the simulation must include \emph{both fermionic and bosonic degrees
of freedom} (matter and gauge fields/particles); It must have a \emph{Relativistic,
Lorentz invariance} (a causal structure); and it must involve \emph{local
gauge invariance}, in order to obtain conservation of charges, and
of course - the required interactions. These requirements are not
trivially met, however, they must be satisfied, as without them,
the theories in question could not be achieved. As we show in this paper, these requirements may be satisfied
and fulfilled in certain configurations of atomic systems.

\subsection{Proposals for Quantum Simulations of High Energy Physics}

So far, only theoretical proposals have been suggested for quantum
simulations of high energy physics.

The first group of works contains condensed matter systems in which
gauge invariance emerges, which may serve as quantum simulators for
Abelian gauge theories. These include, for example, an effective emergence
of a $U(1)$ spin liquid in Pyrochlore \cite{Pyrochlore}; A ring
exchange model using molecular states in optical lattices, yielding
a Coulomb phase $U(1)$ in the limit of no hopping \cite{RingExchange};
The effective emergence of artificial photons in dipolar bosons in
an optical lattice \cite{ArtificialPhotons}; a \emph{digital} quantum
simulation of spin-$\frac{1}{2}$ $U(1)$ gauge theory, as a low
energy theory of Rydberg atoms \cite{Rydberg}; and the emergence
of gauge fields out of time reversal symmetry breaking for spin-$\frac{5}{2}$
in honeycomb optical lattices \cite{Szirmai2011}.

Quantum simulation, using ultracold atoms in an optical lattice, of
continuous $QED$ has been suggested in \cite{Kapit2011}. However,
most of the works so far concentrate on quantum simulations of lattice
gauge theories, which are more suited for quantum simulation using
optical lattices.

Simulations of pure-gauge compact $QED$ ($cQED$, see section \ref{secCQED}) include
a simulation of the Kogut-Susskind Hamiltonian in 2+1 dimensions using
Bose-Einstein condensates in optical lattices \cite{Zohar2011}, as well as a simulation of a truncated, spin-gauge
model, using single atoms in optical lattices \cite{Zohar2012}.

Simulations of $U(1)$ models with matter have been suggested as well
- either for an Abelian link model \cite{Horn1981,Orland1990,wiese1997}
- a 1+1 dimensional simulation of the lattice link Schwinger model
\cite{Banerjee2012}, or a generalization of the spin-gauge approach,
to include dynamical fermions in 2+1 dimensions \cite{Zohar2013}. The above proposals for $cQED$ simulations with
ultracold atoms introduce gauge invariance as an effective symmetry,
by constraining Gauss's law in the Hamiltonian, with a large energy
penalty. Another approach for simulation, utilizing fundamental symmetries
of the ultracold atoms, is proposed in \cite{AngMom}, for 1+1 and 2+1 dimensional $cQED$, with or without
dynamical matter.

Recent proposals for such simulations also utilize other systems -
such as superconducting quantum circuits \cite{SQC,Marcos2014} and trapped ions
\cite{ZollerIons}, both simulating the Abelian link model. A digital
simulation of a $U(1)$ gauge theory is presented in \cite{Tagliacozzo2013},
using Rydberg atoms.

Simulations of discrete lattice gauge theories have been suggested
as well, including discrete Abelian ($\mathbb{Z}_{N}$) and non-Abelian ($D_N$)
models using arrays of Josephson junctions \cite{Doucot2004}, a digital
simulation of $\mathbb{Z}_{2}$ with Rydberg atoms \cite{Tagliacozzo2013}
and an analog simulation of $\mathbb{Z}_{N}$, with an explicit construction
for $\mathbb{Z}_{3}$ in an optical lattice \cite{AngMom}.
A way to truncate the Hilbert space of a $U(1)$ theory in a manner similar
to $\mathbb{Z}_N$ was proposed in \cite{Notarnicola2015}.

Quantum simulations of Non-Abelian gauge theories with continuous
groups ($SU(2)$ Yang Mills models, with possible generalizations to $U(N),SU(N)$) have been proposed too - Analog simulations
with cold atoms utilizing prepotentials \cite{MathurSU2,Mathur2006,Mathur2007,Anishetty}
or rishons in the link model \cite{Brower1999,Brower2004} have been
suggested in \cite{NA} and \cite{Rishon2012}
respectively, where the first utilizes hyperfine angular momentum
conservation to obtain local gauge invariance, and the latter starts
with a larger symmetry group, which must be broken. A digital simulation
of an $SU(2)$ gauge magnet \cite{Orland1990,Orland2012} has been suggested in \cite{TagliacozzoNA},
while a digital quantum simulation of an $SU(2)$ link model with triangular plaquettes,
using superconducting qubits, has
been suggested in \cite{Mezzacapo2015}.
A new way for imposing gauge invariance using dissipation
(Zeno effect), with an explicit construction for a non-Abelian model,
has been given in \cite{Dissipation}.

In general, the dimensions of the analog quantum simulations with optical lattices are
restricted only by experimental and technological considerations. Thus, the above suggestions
for $2+1$ dimensional systems, using optical lattices, should apply, in principle, also to $d+1$ systems,
given, of course, that the required technological challenge is met. The $1+1$ quantum simulations mentioned above
which use optical lattices may be generalized to further dimensions but, mostly, without the magnetic (plaquette) terms,
and thus are simply quantum simulations the strong limit of lattice gauge theories (see section \ref{secLGT} for
an explanation of these terms). For other simulating systems and/or digital simulations, the case may be different,
due to possible dimension-dependent properties of the simulating system.

For a comprehensive review on the recent progress in quantum simulation proposals of link-Rishon models,
the reader should refer to \cite{Wiese2013}.

Throughout this work, $\hbar = c = 1$, and the Einstein summation convention is assumed.

\section{Lattice Gauge Theory: A brief review} \label{secLGT}
Lattice gauge theories \cite{KogutLattice,Kogut1983,Creutz1983,Montvay1997,Smit2002,Rothe2005}
are formulations of gauge theories on a discretized space, or spacetime.
They were originally invented by Wilson, as a tool for the study of the
quark confinement problem \cite{Wilson}. It is an important tool
in high energy physics, and especially in $QCD$ (Quantum Chromodynamics
- the theory of the strong interactions) as it allows performing nonperturbative
calculations (numerically, using Monte-Carlo methods) and thus provides
insights into the perturbatively-inaccessible regions of $QCD$, due
to the running coupling.

Besides addressing the confinement of quarks (or charges in analogous
Abelian theories, as will be later discussed), lattice gauge theories
have enabled the computation of other important quantities in $QCD$,
such as the hadronic structure and spectrum, as may be read in the reviews \cite{Kogut1983,Fukushima2011,FLAG2013} and many others.

In the following review of lattice gauge theory, we follow the Hamiltonian (canonical),
Kogut-Susskind formulation of lattice gauge theories \cite{KogutSusskind,KogutLattice,Kogut1983}
(rather than the more "conventional" Euclidean approach). This is since the Hamiltonian language is much
more natural in the context of atomic and optical physics, used for the simulating systems.

\subsection{Hamiltonian Formulation}

\begin{figure}[t]
\begin{centering}
\includegraphics[scale=0.5]{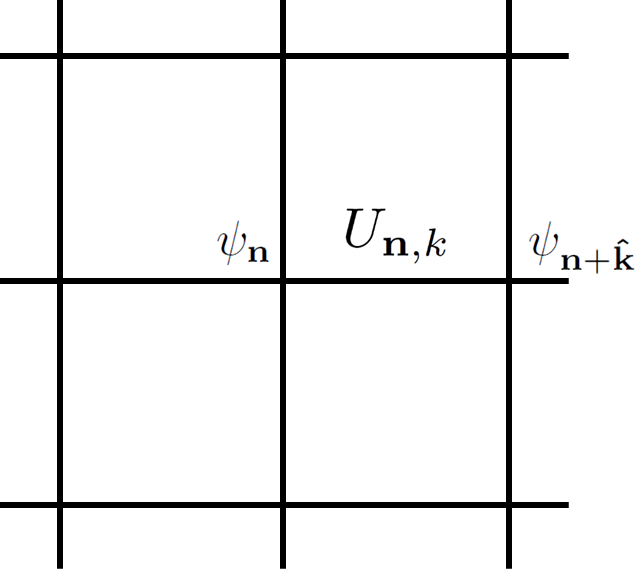}
\par\end{centering}
\caption{The lattice structure. The gauge degrees of freedom occupy the links, whereas the fermionic matter - the vertices.}
\label{fig1}
\end{figure}

\subsubsection{Hopping fermions and global gauge symmetry.}

Consider fermions on a $d$ dimensional spatial lattice, with nearest-neighbor
hopping. The Hamiltonian takes the form
\begin{equation}
H=\underset{\mathbf{n}}{\sum}M_{\mathbf{n}}\psi_{\mathbf{n}}^{\dagger}\psi_{\mathbf{n}}+\epsilon\underset{\mathbf{n},k}{\sum}\left(\psi_{\mathbf{n}}^{\dagger}\psi_{\mathbf{n+\mathbf{\hat{k}}}}+h.c.\right)\label{eq:Hhop}
\end{equation}
where $\mathbf{n}\in\mathbb{Z}^{d}$ label the lattice's vertices
and $k\in\left\{ 1,...,d\right\} $ labels the lattice's directions;
$\mathbf{\hat{k}}$ is a unit vector (where the lattice spacing is
$a=1$) in the $k$th direction. The Hamiltonian consists of local
{}``mass'' terms and hopping terms (see figure \ref{fig1}).

$\psi_{\mathbf{n}}$ are spinors of some unitary gauge group $G$,
in some representation $r$
\footnote{Note that we only consider here the \emph{gauge} degrees of freedom;
Each component of these gauge spinors may also be a spinor in terms
of the spin or flavor degrees of freedom, but we disregard it here.
}. For example, $\psi_{\mathbf{n}}$ are
merely fermionic annihilation operators in the case  $G=U\left(1\right)$. For $G=SU(N)$, in the fundamental
representation, these are $N$-component spinors, containing $N$
annihilation operators:
\begin{equation}
\psi_{\mathbf{n}}=\left(\begin{array}{c}
\psi_{\mathbf{n},1}\\
\psi_{\mathbf{n},2}\\
\vdots\\
\psi_{\mathbf{n},N}
\end{array}\right)
\end{equation}
where in Hamiltonian (\ref{eq:Hhop}), and in what follows, summation
on the \emph{group} indices is assumed, e.g.
\begin{equation}
\psi_{\mathbf{n}}^{\dagger}\psi_{\mathbf{n}}\equiv\underset{i}{\sum}\psi_{\mathbf{n},i}^{\dagger}\psi_{\mathbf{n},i}
\end{equation}

The Hamiltonian (\ref{eq:Hhop}) is invariant under \emph{global gauge
transformations} of the gauge group $G$. That means, that if we pick
some group element $V\in G$ (which is unitary) and perform the transformation
\begin{equation}
\psi_{\mathbf{n}}\longrightarrow V\psi_{\mathbf{n}}\quad;\quad\psi_{\mathbf{n}}^{\dagger}\longrightarrow\psi_{\mathbf{n}}^{\dagger}V^{\dagger}\label{eq:glotran}
\end{equation}
the Hamiltonian is left intact
\footnote{The group element $V$ in (\ref{eq:glotran}) is, of course, written
in the appropriate representation (the same as the spinors). However,
note that the equation is representation independent. When a representation
is chosen it must be identical for the spinor and the group element.
}. This corresponds to a conservation of the total number of fermions,
\begin{equation}
N_{tot}=\underset{\mathbf{n}}{\sum}\psi_{\mathbf{n}}^{\dagger}\psi_{\mathbf{n}}
\end{equation}
which is obvious, as all the terms of the Hamiltonian annihilate one
fermion and create another one.

\subsubsection{\label{sub:Local-Gauge-Symmetry}Local Gauge Symmetry.}

Next, let us lift the symmetry to be local, by generalizing the transformation
to
\begin{equation}
\psi_{\mathbf{n}}\longrightarrow V_{\mathbf{n}}\psi_{\mathbf{n}}\quad;\quad\psi_{\mathbf{n}}^{\dagger}\longrightarrow\psi_{\mathbf{n}}^{\dagger}V_{\mathbf{n}}^{\dagger}
\end{equation}
- i.e., each vertex is assigned \emph{locally} a unitary element of
the gauge group. The mass terms of (\ref{eq:Hhop}) are local and
hence remain intact, but the hopping terms vary under the transformation,
for example
\begin{equation}
\psi_{\mathbf{n}}^{\dagger}\psi_{\mathbf{n+\hat{k}}}\longrightarrow\psi_{\mathbf{n}}^{\dagger}V_{\mathbf{n}}^{\dagger}V_{\mathbf{n+\mathbf{\hat{k}}}}\psi_{\mathbf{n+\mathbf{\hat{k}}}}
\end{equation}

The gauge symmetry will be restored if we introduce a connection $U_{\mathbf{n},k}$,
located on the link emanating from the vertex $\mathbf{n}$ in the
$k$th direction. $U_{\mathbf{n},k}$ is an element
of the gauge group in the appropriate representation (corresponding
to the spinors to which it is coupled), a unitary operator
\footnote{The unitarity has many advantages and is used in the context of conventional, Wilsonian minimally-coupled
lattice gauge theories. However, the transformation (symmetry) properties may
be satisfied if the operator is not unitary as well, as we do in our
simulations \cite{Zohar2012,Zohar2013,AngMom,NA}.
} undergoing the gauge transformation as
\begin{equation}
U_{\mathbf{n},k}\longrightarrow V_{\mathbf{n}}U_{\mathbf{n},k}V_{\mathbf{n+\mathbf{\hat{k}}}}^{\dagger}
\end{equation}
then, if the Hamiltonian is changed to
\begin{equation}
H_{GM}=\underset{\mathbf{n}}{\sum}M_{\mathbf{n}}\psi_{\mathbf{n}}^{\dagger}\psi_{\mathbf{n}}+\epsilon\underset{\mathbf{n},k}{\sum}\left(\psi_{\mathbf{n}}^{\dagger}U_{\mathbf{n},k}\psi_{\mathbf{n+\mathbf{\hat{k}}}}+h.c.\right)\label{eq:HGM}
\end{equation}
the gauge invariance is, indeed, restored, and the gauge transformation
leaves it intact. The name $H_{GM}$ stands for the gauge-matter interactions
described by it.

\begin{figure}[t]
\begin{centering}
\includegraphics[scale=0.7]{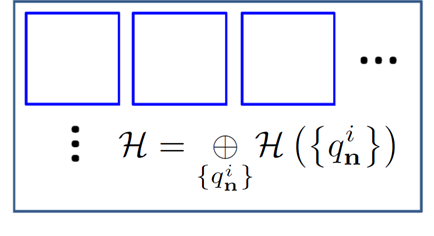}
\par\end{centering}

\caption{The physical Hilbert space of a gauge invariant theory is divided
into separate sectors, corresponding to different static charge configurations
(\ref{eq:hilbdec}), represented by boxes. The gauge invariant dynamics does not mix among
them (\ref{eq:hilbnomix}). These are two consequences of gauge invariance
(\ref{eq:gausscomm}).\label{fig:The-physical-Hilbert}}

\end{figure}

We denote the local gauge transformation generators by $G_{\mathbf{n}}^{i}$.
Due to the gauge symmetry, they commute with the Hamiltonian,
\begin{equation}
\left[H,G_{\mathbf{n}}^{i}\right]=0\label{eq:gausscomm}
\end{equation}
and thus the physical Hilbert space $\mathcal{H}$ is also gauge-invariant,
i.e. divided into sectors of eigenvalues of $G_{\mathbf{n}}^{i}$
(see figure \ref{fig:The-physical-Hilbert}):
\begin{equation}
\mathcal{H}=\underset{\left\{ q_{\mathbf{n}}^{i}\right\} }{\oplus}\mathcal{H}\left(\left\{ q_{\mathbf{n}}^{i}\right\} \right)\label{eq:hilbdec}
\end{equation}
 Each such sector $\mathcal{H}\left(\left\{ q_{\mathbf{n}}^{i}\right\} \right)$
has a set of eigenvalues $\left\{ q_{\mathbf{n}}^{i}\right\} $, called
\emph{static charges}, such that for every $\left|\psi\left(\left\{ q_{\mathbf{n}}^{i}\right\} \right)\right\rangle \in\mathcal{H}\left(\left\{ q_{\mathbf{n}}^{i}\right\} \right)$,
\begin{equation}
G_{\mathbf{n}}^{i}\left|\psi\left(\left\{ q_{\mathbf{n}}^{i}\right\} \right)\right\rangle =q_{\mathbf{n}}^{i}\left|\psi\left(\left\{ q_{\mathbf{n}}^{i}\right\} \right)\right\rangle \label{eq:gausslaw}
\end{equation}
Note that as the generators do not commute for a non-Abelian group (this
shall be clear in the following subsection, where they are explicitly
defined).  The static charges, thus, can only be defined by the eigenvalues
of a subset of some commuting operators.
The above equation is the generalized \emph{Gauss's Law, }and the
dynamics does not mix these sectors, i.e.
\begin{equation}
\left\langle \psi\left(\left\{ p_{\mathbf{n}}^{i}\right\} \right)\right|H\left|\psi\left(\left\{ q_{\mathbf{n}}^{i}\right\} \right)\right\rangle \propto\underset{\mathbf{n}}{\prod}\delta_{p_{\mathbf{n}}^{i},q_{\mathbf{n}}^{i}}\label{eq:hilbnomix}
\end{equation}

\subsubsection{Dynamical Gauge Fields.}

We shall identify the connections $U_{\mathbf{n},\mathbf{\hat{k}}}$
with the gauge group elements, i.e. introduce the
group parameters $\theta_{\mathbf{\mathbf{n},}k}^{i}$ - Hermitian
operators, such that in a given representation $r$, with the matrix
representation of the group generators $T_{i}^{\left(r\right)}$,
\begin{equation}
U_{\mathbf{n},k}^{\left(r\right)}=e^{i\theta_{\mathbf{\mathbf{n},}k}^{i}T_{i}^{\left(r\right)}}
\end{equation}
where a summation on the group indices $i$ is implicit. We shall
assume next working in the fundamental representation (generalizations
to other representations are straightforward), and disregard the representation
indices in this case: $U_{\mathbf{n},k}=e^{i\theta_{\mathbf{\mathbf{n},}k}^{i}T_{i}}$.

One could then define operators which are conjugate to the group parameters,
to serve as generalized momenta. In case of the Abelian group $U\left(1\right)$,
where $U_{\mathbf{n},k}=e^{i\theta_{\mathbf{\mathbf{n},}k}}$, one
could define the conjugate momentum to be proportional simply to $\dot{\theta}_{\mathbf{\mathbf{n},\mathbf{\hat{k}}}}$.
But generally, when $G$ may be non-Abelian, one must pay attention
of the non-commutativity of the group in doing that. Thus, we could
reformulate the Abelian case and define the conjugate momentum there
as
\begin{equation}
L_{\mathbf{\mathbf{n},}k}=-i\dot{U}_{\mathbf{n},k}U_{\mathbf{n},k}^{\dagger}=-iU_{\mathbf{n},k}^{\dagger}\dot{U}_{\mathbf{n},k}
\end{equation}
The differential representation would be
\begin{equation}
L_{\mathbf{\mathbf{n},}k}=-i\frac{\partial}{\partial\theta_{\mathbf{\mathbf{n},}k}}
\end{equation}
from which the canonical commutation relation
\begin{equation}
\left[\theta_{\mathbf{\mathbf{n},}k},L_{\mathbf{\mathbf{n},}k}\right]=i\label{eq:ELcomm}
\end{equation}
results, and from which, as it is the generator of angular translations,
\begin{equation}
\left[L_{\mathbf{\mathbf{n},}k},U_{\mathbf{\mathbf{n},}k}\right]=U_{\mathbf{\mathbf{n},}k}
\end{equation}
Thus a reasonable choice for the dynamic part of the gauge field would
be
\begin{equation}
H_{E}^{ab}\propto\underset{\mathbf{n},k}{\sum}L_{\mathbf{\mathbf{n},}k}^{2}
\end{equation}
in agreement with the Abelian version of the Kogut-Susskind Hamiltonian (see below).

Let us now generalize it for non-Abelian groups \cite{Polyakov1987}.
Due to the non-Abelian nature of the group, one could define two sets
of such operators, left and right -
\begin{equation}
L_{\mathbf{n},k}=-i\dot{U}_{\mathbf{n},k}U_{\mathbf{n},k}^{\dagger}\quad;\quad R_{\mathbf{n},k}=-iU_{\mathbf{n},k}^{\dagger}\dot{U}_{\mathbf{n},k}\label{eq:leftright}
\end{equation}
(for an Abelian group $L=R$, but this does not hold in general).
One can expand these operators in terms of the group's representation
matrices,
\begin{equation}
L_{\mathbf{n},k}=L_{\mathbf{n},k}^{i}T_{i}\quad;\quad R_{\mathbf{n},k}=R_{\mathbf{n},k}^{i}T_{i}
\end{equation}
and then construct the differential forms of the operators $\left\{ L_{\mathbf{n},k}^{a}\right\} ,\left\{ R_{\mathbf{n},k}^{a}\right\} $.
These operators are called \emph{left and right} generators of the
group, since they satisfy the group's algebra
\footnote{Note that the right and left operators may be defined in other conventions
as well, resulting in different signs in the group's algebra.
},
\begin{equation}
\left[L_{\mathbf{n},k}^{i},U_{\mathbf{n},k}\right]=T_{i}U_{\mathbf{n},k}\label{eq:alg1}
\end{equation}
\begin{equation}
\left[R_{\mathbf{n},k}^{i},U_{\mathbf{n},k}\right]=U_{\mathbf{n},k}T_{i}\label{eq:alg2}
\end{equation}
\begin{equation}
\left[L_{\mathbf{n},k}^{i},L_{\mathbf{n},k}^{j}\right]=-if_{ijl}L_{\mathbf{n},k}^{l}\label{eq:alg3}
\end{equation}
\begin{equation}
\left[R_{\mathbf{n},k}^{i},R_{\mathbf{n},k}^{h}\right]=if_{ijl}L_{\mathbf{n},k}^{l}\label{eq:alg4}
\end{equation}
\begin{equation}
\left[L_{\mathbf{n},k}^{i},R_{\mathbf{n},k}^{j}\right]=0\label{eq:alg5}
\end{equation}
where $\left\{ T_{i}\right\} $ are the group's generators and $f_{ijl}$
are the group's structure constants
\footnote{In a general representation $r$, these equations would be generalized
in a straightforward manner, e.g. $\left[L_{\mathbf{n},k}^{i},U_{\mathbf{n},k}^{\left(r\right)}\right]=T_{i}^{\left(r\right)}U_{\mathbf{n},k}^{\left(r\right)}$
, where $T_{i}^{\left(r\right)}$ stands for the matrix representation
of the generator in $r$, and $U_{\mathbf{n},k}^{\left(r\right)}$
stands for the matrix representation of the group element. If one
wishes to explicitly write the group's indices, the equations will
look like, for example, $\left[L_{\mathbf{n},k}^{i},\left(U_{\mathbf{n},k}\right)_{jl}\right]=\left(T_{i}U_{\mathbf{n},k}\right)_{jl}$
}. Generators and group elements from different links commute, of course.

Note that due to the definition (\ref{eq:leftright}), one could prove
that
\begin{equation}
R_{\mathbf{n},k}=U_{\mathbf{n},k}^{\dagger}L_{\mathbf{n},k}U_{\mathbf{n},k}
\end{equation}
and thus if we calculate the trace in group space we find out that
\begin{equation}
\mathrm{tr}\left(R_{\mathbf{n},k}^{2}\right)=\mathrm{tr}\left(L_{\mathbf{n},k}^{2}\right)
\end{equation}
This can be reformulated using the generators,
\begin{equation}
\underset{i}{\sum}\left(R_{\mathbf{n},k}^{i}\right)^{2}=\underset{i}{\sum}\left(L_{\mathbf{n},k}^{i}\right)^{2}\equiv\mathbf{L}_{\mathbf{\mathbf{n},}k}^{2}
\end{equation}
giving us the electric part of the non-Abelian Kogut-Susskind Hamiltonian,
\begin{equation}
H_{E}\propto\underset{\mathbf{n},k}{\sum}\mathbf{L}_{\mathbf{\mathbf{n},}k}^{2}
\label{HE}
\end{equation}
We can refer to $\left\{ L_{\mathbf{n},\mathbf{\hat{k}}}^{i}\right\} ,\left\{ R_{\mathbf{n},\mathbf{\hat{k}}}^{i}\right\} $
as the left and right electric fields, conjugate (in some sense) to
the vector potential. Their difference along a link corresponds to
the charge carried by it: no charge in the Abelian case, and the color
charge in the $SU\left(N\right)$ case.

The gauge transformation generators are then given by \cite{KogutSusskind}
\begin{equation}
G_{\mathbf{n}}^{i}=\underset{k}{\sum}\left(L_{\mathbf{n},k}^{i}-R_{\mathbf{n}-\mathbf{\hat{k}},k}^{i}\right)-Q_{\mathbf{n}}^{i}\label{eq:Gaugegen}
\end{equation}
where $Q_{\mathbf{n}}^{a}$ is the \emph{dynamical }charge, which
will be introduced later.

The Abelian version of the generator is merely
\begin{align}
\begin{aligned}G_{\mathbf{n}} & = & \underset{k}{\sum}\left(L_{\mathbf{n},k}-L_{\mathbf{n}-\mathbf{\hat{k}},k}\right)-Q_{\mathbf{n}}\\
 & \equiv & \mathrm{div}_{\mathbf{n}}L_{\mathbf{n},k}-Q_{\mathbf{n}}
\end{aligned}
\label{eq:gaugeab}
\end{align}
in which the discrete divergence is defined, and from which the relation
to Gauss's law is obvious.

\begin{figure}[t]

\begin{centering}
\includegraphics[scale=0.4]{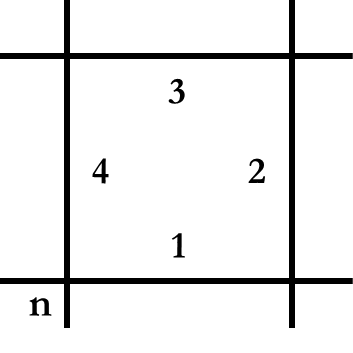}\caption{The convention of numbering the links in plaquette interactions (see
equations (\ref{eq:HBBB}),(\ref{eq:HBBBB})).\label{fig:The-convention-of}}

\par\end{centering}

\end{figure}

One would also like to consider a gauge-field self interaction. Gauge
invariance forces us to consider only group traces of products of
group elements along close paths. The smallest such interactions are
the so-called \emph{plaquette} terms, involving unit squares of the
lattice, from which we obtain the magnetic part of the Kogut-Susskind
Hamiltonian \cite{KogutSusskind}
\begin{equation}
H_{B}\propto\underset{plaquettes}{\sum}\left(\mathrm{tr}\left(U_{1}U_{2}U_{3}^{\dagger}U_{4}^{\dagger}\right)+h.c.\right)\label{eq:HBBB}
\end{equation}
with the group elements usually in the fundamental representation
(even if there are charges in other representation which requires
the use of other representations when coupling them to the gauge field)
which simplifies in the Abelian case to \cite{KogutLattice}
\begin{equation}
H_{B}^{ab}\propto\underset{plaquettes}{\sum}\cos\left(\theta_{1}+\theta_{2}-\theta_{3}-\theta_{4}\right)\label{eq:HBBBB}
\end{equation}
where the numbers in both the equations are according to the plaquette
convention presented in figure \ref{fig:The-convention-of}. Note
that this agrees with the terms of the Wilson action \cite{Wilson},
and thus with the {}``usual'' procedure of obtaining the Hamiltonian (by a Legendre transformation of the Lagrangian).

We have obtained the terms of the Kogut-Susskind Hamiltonian only
up to some proportion factors. One can find them by taking the (classical)
continuum limit. Doing that (or using a Legendre transformation of the Lagrangian)
one obtains the Kogut-Susskind Hamiltonian
\begin{equation}
H_{KS}=\frac{1}{2}g^{2}\underset{\mathbf{n},k}{\sum}\mathbf{L}_{\mathbf{\mathbf{n},}k}^{2}-\frac{1}{2g^{2}}\underset{plaquettes}{\sum}\left(\mathrm{tr}\left(U_{1}U_{2}U_{3}^{\dagger}U_{4}^{\dagger}\right)+h.c.\right)\label{eq:KS_Ham}
\end{equation}
where $g$ is the coupling constant.

\subsection{Abelian Theories}
\subsubsection{Compact QED (cQED)} \label{secCQED}
The first theory we shall consider is Compact Quantum Electrodynamics
- $cQED$ \cite{KogutLattice}. This theory's continuum limit is Quantum
Electrodynamics ($QED$). Its lattice version is compact, involving
angular variables.

This is an Abelian, continuous gauge theory, with the gauge group
$U\left(1\right)$. Thus, the group elements are simply unitary phases
\begin{equation}
U_{\mathbf{n},k}=e^{i\theta_{\mathbf{\mathbf{n},}k}}
\end{equation}
as we have already discussed, $\theta_{\mathbf{\mathbf{n},}k}$ are
analogous to the vector-potential, and the conjugate momenta are $L_{\mathbf{\mathbf{n},}k}=-i\frac{\partial}{\partial\theta_{\mathbf{\mathbf{n},}k}}$
- $U(1)$ (planar) angular momentum operators. The canonical commutation
relation $\left[\theta_{\mathbf{\mathbf{n},}k},L_{\mathbf{\mathbf{n},}k}\right]=i$
ensures that $U_{\mathbf{\mathbf{n},}k}$ is a raising operator of
the electric field $\left[L_{\mathbf{\mathbf{n},}k},U_{\mathbf{\mathbf{n},}k}\right]=U_{\mathbf{\mathbf{n},}k}$,
and $U_{\mathbf{\mathbf{n},}k}^{\dagger}$ is the lowering operator,
both are unitary.

The Kogut-Susskind Hamiltonian takes the form
\begin{equation}
H_{KS}=\frac{g^{2}}{2}\underset{\mathbf{n},k}{\sum}L_{\mathbf{\mathbf{n},}k}^{2}-\frac{1}{g^{2}}\underset{plaquettes}{\sum}\cos\left(\theta_{1}+\theta_{2}-\theta_{3}-\theta_{4}\right)\label{eq:abKS}
\end{equation}
where the indices inside the plaquette term are with respect to the convention
of figure \ref{fig:The-convention-of}. Also note that the argument
of the cosine corresponds to the lattice curl, manifesting that
the continuum limit of this term gives rise to the magnetic energy.

Let us consider the local Hilbert space on each link. One could work
in the basis of electric (flux) eigenstates, which is a good basis
in the strong coupling limit $g^{2}\gg1$, as shall be discussed when we consider confinement.
The eigenstates of $L$ on a single link satisfy
\begin{equation}
L\left|m\right\rangle =m\left|m\right\rangle
\end{equation}
where $m$ takes any integer value, either positive or negative. In
coordinate (angle) representation, the wavefunctions are
\begin{equation}
\phi_{m}\left(\theta\right)=\left\langle \theta|m\right\rangle =\frac{1}{\sqrt{2\pi}}e^{im\theta}
\end{equation}

The Abelian Gauss's law takes the form
\begin{align}
\begin{aligned}G_{\mathbf{n}} & = & \underset{k}{\sum}\left(L_{\mathbf{n},k}-L_{\mathbf{n}-\mathbf{\hat{k}},k}\right)-Q_{\mathbf{n}}\\
 & \equiv & \mathrm{div}_{\mathbf{n}}L_{\mathbf{n},k}-Q_{\mathbf{n}}
\end{aligned}
\end{align}
and thus, in this basis, the gauge-invariant states satisfy (without
dynamical matter)
\begin{equation}
G_{\mathbf{n}}\left|\phi\right\rangle =\underset{k}{\sum}\left(m_{\mathbf{n},k}-m_{\mathbf{n}-\mathbf{\hat{k}},k}\right)\left|\phi\right\rangle =q_{\mathbf{n}}\left|\phi\right\rangle
\end{equation}
where $q_{\mathbf{n}}$ is the (C-number) static charge.

What are the interactions?
\begin{enumerate}
\item The plaquette interaction, in this language, changes the electric
flux on a plaquette, such that in two links the flux increases by
a single unit, and in the other two it is lowered. This is done only in two possible
orientations, which are the only possibilities which leave
the eigenvalue of $G_{\mathbf{n}}$ intact in all the four vertices
of each plaquette - thus, this is the {}``physical interpretation''
of the commutativity of the plaquette interactions with Gauss's law.
\item If we wish to introduce dynamical charges, staggered \cite{KogutSusskind,Susskind1977}, for example
\footnote{Note that staggered fermions are not the only possible recipe for discretization of fermionic fields,
required by the doubling problem \cite{Susskind1977,Karsten1981,Kogut1983,Creutz1983,Montvay1997,Smit2002,Rothe2005}. The different approaches, which are out of the scope of this paper,
vary in both quantitative and qualitative details, and have their own advantages and disadvantages. Yet, when a quantum simulator is designed, one should consider which type of lattice fermions
to use. We mostly use the staggered fermions formulation, but other approaches should be possible as well.}, with
a Hamiltonian of the form
\begin{equation}
H_{M}=\epsilon\underset{\mathbf{n},k}{\sum}\left(\psi_{\mathbf{n}}^{\dagger}e^{i\theta_{\mathbf{\mathbf{n},}k}}\psi_{\mathbf{n+\hat{k}}}+H.c.\right)+m\underset{\mathbf{n}}{\sum}\left(-1\right)^{\underset{k}{\sum}n_{k}}\psi_{\mathbf{n}}^{\dagger}\psi_{\mathbf{n}}
\end{equation}
where $\left\{n_k\right\}$ are the indices of the vertex $\mathbf{n}$.
we may interpret the gauge invariant interactions as ones which raise
the dynamical charge on one edge of the link, lower it on the other
side, and raise/lower the electric flux on the link in a way that
Gauss's law, now including the dynamical charges, will hold in both
the edges.
In this staggered representation, even vertices represent {}``particles'', with a positive mass $m$,
and odd ones - {}``holes'', with a negative mass $-m$. In fact,
the negative vertices may be interpreted as the Dirac sea: suppose
we shift the energy by a constant, and measure the mass of the odd
vertices with respect to $-m$. Then an occupied odd vertex corresponds
to a state with zero mass - no particle at all, and a vacant odd vertex
corresponds to an anti-particle with mass $m$. This corresponds to
the canonical transformation of second quantized Dirac fields, where
the holes' creation operators are replaced with anti-particles' annihilation
operators.
We can also define the charges of the particles: the particle would
have charge 1 (in units of the fundamental charge, of coupling constant
$g$), and the anti-particle - charge -1. This means, in the language
of fermionic operators, that
\begin{equation}
Q_{n}=\psi_{n}^{\dagger}\psi_{n}-\frac{1}{2}\left(1-\left(-1\right)^{\underset{k}{\sum}n_{k}}\right)
\end{equation}
and this completes the particle--anti-particle picture.

\end{enumerate}

These two possible interactions may be understood as the meaning of
\emph{local gauge invariance}. Note that the only two possibilities
of contracting gauge invariant operators are traces of products of
group elements along closed loops, or along open loops, bounded by
two fermionic operators on the edges. These two interactions are the
shortest such interactions possible, and thus the most local ones.
The same considerations apply as well to other lattice gauge theories.

\subsubsection{$\mathbb{Z}_N$.}

Another Abelian gauge theory of relevance is $\mathbb{Z}_{N}$ . This is
a \emph{discrete} group, and thus this is not a Yang-Mills theory,
and its structure is a little different. Nevertheless one can use
this group to formulate a lattice gauge theory and observe interesting
physics. This group is highly relevant for the confinement in $QCD$,
as $\mathbb{Z}_{3}$ is the center of this group, and the center is
responsible for confinement (large-distance phenomena) \cite{Hooft1978}.

On every link of the lattice, we define two unitary operators, $P$
and $Q$ \cite{Horn1979}, such that
\begin{equation}
P^{\dagger}P=Q^{\dagger}Q=1
\end{equation}
which satisfy the $\mathbb{Z}_{N}$ algebra,
\begin{equation}
P^{N}=Q^{N}=1\quad;\quad P^{\dagger}QP=e^{i\delta}Q
\end{equation}
where $\delta=\frac{2\pi}{N}$.

Let us define the eigenstates of $P$ as
\begin{equation}
P\left|m\right\rangle =e^{im\delta}\left|m\right\rangle
\end{equation}
there are $N$ such states,
\begin{equation}
m\in\left\{ -\frac{N-1}{2},...,\frac{N-1}{2}\right\}
\end{equation}
for an odd $N$ (the generalization for an even $N$ is straightforward).
$Q$ is a unitary ladder operator,
\begin{equation}
Q\left|m\right\rangle =\left|m-1\right\rangle
\end{equation}
with the cyclic property
\begin{equation}
Q\left|-\frac{N-1}{2}\right\rangle =\left|\frac{N-1}{2}\right\rangle
\end{equation}
alternatively, one may use the eigenstates of $Q$, and then $P$
will be a unitary raising operator, cyclic as well.

One can define Hermitian operators $E,A$ such that
\begin{equation}
P=e^{i\delta E}\qquad Q=e^{iA}
\end{equation}
and in the $N\longrightarrow\infty$ limit they will correspond to
$cQED$'s conjugate electric field and vector potential.

The Hamiltonian of this theory takes the form \cite{Horn1979}
\begin{equation}
H=-\frac{\lambda}{2}\underset{\mathbf{n},k}{\sum}\left(P_{\mathbf{n},k}+P_{\mathbf{n},k}^{\dagger}\right)-\frac{1}{2}\underset{plaq.}{\sum}\left(Q_{1}Q_{2}Q_{3}^{\dagger}Q_{4}^{\dagger}+H.c.\right)
\end{equation}
with local terms and plaquette interactions, which tend to the Abelian
Kogut-Susskind Hamiltonian (\ref{eq:abKS}) as $N\longrightarrow\infty$.
The plaquette indexing convention is, again, according to figure \ref{fig:The-convention-of}.

One can define static modular charges the vertices
\begin{equation}
q_{\mathbf{n}}=e^{-i\delta m}
\end{equation}
and then the Gauss's law is given by
\begin{equation}
G_{\mathbf{n}}\left|\phi\right\rangle =q_{\mathbf{n}}\left|\phi\right\rangle
\end{equation}
where
\begin{equation}
G_{\mathbf{n}}=\underset{l+}{\prod}P_{l+}^{\dagger}\underset{l-}{\prod}P_{l-}
\end{equation}
with $l_{+}$ being links which start from $\mathbf{n}$ (positive
links) and $l_{-}$ - ending there (negative ones).

A description of gauge theories with finite groups (both Abelian and non-Abelian),
with dynamical, fermionic matter as well, is found in \cite{PRD}.

\subsection{Non-Abelian Yang-Mills Theories: $SU(2)$ as an example}

As an example of a continuous Non-Abelian gauge theory, we examine
next the $SU(2)$ gauge theory \cite{KogutSusskind,Kogut1983}. $SU(2)$
(the rotation group) has three generators, with the structure constants
\begin{equation}
f_{ijk}=\epsilon_{ijk}
\end{equation}
which implies for the left and right algebras
\begin{equation}
\left[L_{\mathbf{n},k}^{i},L_{\mathbf{n},k}^{j}\right]=-i\epsilon_{ijl}L_{\mathbf{n},k}^{l}\label{eq:alg3-1}
\end{equation}
\begin{equation}
\left[R_{\mathbf{n},k}^{i},R_{\mathbf{n},k}^{j}\right]=i\epsilon_{ijl}R_{\mathbf{n},k}^{l}\label{eq:alg4-1}
\end{equation}

The pure-gauge Hamiltonian is (\ref{eq:KS_Ham}). One may choose the
representation of the $U$ matrices, and let us do that in the fundamental
representation of $SU\left(2\right)$ ($j=\frac{1}{2}$). In this
representation we work with the Pauli matrices,
\begin{equation}
T^{i}=\frac{1}{2}\sigma^{i}
\end{equation}
and thus
\begin{equation}
U_{\mathbf{n},k}=e^{\frac{i}{2}\theta_{\mathbf{\mathbf{n},}k}^{i}\sigma_{i}}
\end{equation}
- which are $2\times2$ matrices of operators.
The reader may find an explicit expression for this operators, using the formulation \cite{PRD},
when we discuss the quantum simulation of this model.

The links' local Hilbert space, in the flux basis, is now described
using three quantum numbers $\left|jmm'\right\rangle $. These are
eigenstates of the operators
\begin{equation}
\mathbf{L}^{2}\left|jmm'\right\rangle =\mathbf{R}^{2}\left|jmm'\right\rangle =j\left(j+1\right)\left|jmm'\right\rangle \label{eq:SU2casimir}
\end{equation}
\begin{equation}
L_{z}\left|jmm'\right\rangle =m\left|jmm'\right\rangle
\end{equation}
\begin{equation}
R_{z}\left|jmm'\right\rangle =m'\left|jmm'\right\rangle
\end{equation}

A mechanical interpretation of this Hilbert space may be a rigid body,
described in both body and space coordinates. In both of them the
total angular momentum is equal, but in these two frames of reference
the components of the angular momentum will be measured differently,
as they form two separate, commuting algebras - the body and the space
algebras, corresponding to the two algebras we have here \cite{KogutSusskind,Landau1981}.

By acting with the group elements, which are rotation matrices, on
the $\left|JMM'\right\rangle $ states, an amount of angular momentum
is given to the state (or taken from it). Unlike in the $U(1)$ case,
where the angular momentum is described by a single quantum number,
here the representation is also a dynamical quantity: for example,
acting with $U_{\mathbf{n},k}$ in the fundamental representation
on a $\left|JMM'\right\rangle $ state means adding a $\frac{1}{2}$
quanta of angular momentum to the state, resulting in a superposition
of $J+\frac{1}{2}$ and $J-\frac{1}{2}$ states (ordinary addition
of angular momentum). Quantitatively this is expressed by \cite{Rose1995}
\begin{equation}
U_{mm'}\left|JMM'\right\rangle =C_{+}\left|J+\frac{1}{2},M+m,M'+m'\right\rangle +C_{-}\left|J-\frac{1}{2},M+m,M'+m'\right\rangle
\end{equation}
where $C_{\pm}$ depend on the appropriate Clebsch-Gordan coefficients
\cite{Rose1995,Edmonds1996}. Thus the representation of the state,
the quantum number $J$, is a dynamical observable here, unlike in
continuum field theories, where each field has a constant, well-defined, static
representation: this is due to the fact that lattice gauge fields are not
Fock-space bosons \cite{PRD}.

If one wishes to include dynamical fermions as well, the interaction
with which shall be described by a Hamiltonian of the form (\ref{eq:HGM}),
where $\psi_{\mathbf{n}}$ are \emph{group spinors. }Thus, if we pick
$U$ to be in the fundamental representation, this will also be the
case for the spinors, which will contain now two components - each
Lorentz component will be such a spinor. With staggered fermions we
shall have such a single color spinor on each vertex. In this case
the charges will simply be \cite{KogutSusskind,Kogut1983}
\begin{equation}
Q_{\mathbf{n}}^{a}=\frac{1}{2}\underset{ij}{\sum}\psi_{\mathbf{n},i}^{\dagger}\left(\sigma^{a}\right)_{i,j}\psi_{\mathbf{n},j}
\end{equation}
satisfying the group's algebra. Note that the {}``Dirac sea'' state
can be obtained here as well (as a no-interactions vacuum). Again,
we measure the mass of each particle with respect to $-m$, and a
fully occupied vertex, due to the fermionic nature of the matter particles,
corresponds to zero charge. Thus the Dirac sea is the case with empty
even vertices and fully occupied (by two fermions) odd vertices \cite{Hamer1977}.

The description of other $SU\left(N\right)$ groups is similar; Changing
the group means changing the local Hilbert spaces, but the Hamiltonian
has the same form. A description of Hamiltonian lattice compact $QCD$
($SU\left(3\right)$) is given in \cite{Kogut2004}.

\subsection{Physical content of the models}

Gauge theories in general, including ones on the lattice, exhibit an interesting phase structure.
These include, for example, $QCD$'s phases of color-superconductivity
or quark-gluon plasma \cite{McLerran1986,Kogut2004,Fukushima2011}, whose treatment using the usual
Monte-Carlo methods utilized in lattice gauge theory is rather problematic, due to the  computationally
hard sign problem \cite{Troyer2005} that limits calculations in regimes
with a finite chemical potential for the fermions.

Another interesting and important property of gauge theories is the existence of confining phases,
dependent on the coupling constant and other parameters. We shall review quark confinement in pure gauge theories in detail below.

\subsubsection{Quark Confinement.}
Free quarks are not found in nature, but rather bind together and
form hadrons, such as baryons and mesons. The reason to that is the
physical phenomenon of quark confinement, manifested in $QCD$ but
also in other gauge theories: in large distances, or low energies,
the static potential between quarks is linear in the distance, and
thus it is energetically unfavorable (and practically impractical)
to separate bound quarks. It is a highly intriguing concept, forming
a fundamental aspect in the understanding of the Hadronic spectrum,
involving nontrivial low energy physics, which cannot be addressed
perturbatively.

Schwinger showed, by exactly solving 1+1 dimensional $QED$ with massless
charges \cite{Schwinger1962I,Schwinger1962}, that vector mesons can
be massive if massless fermions are screened by the vacuum polarization,
obtaining massive photons. Further studies \cite{Lowenstein1971,Casher1974}
have shown that only massive photons, no electrons, exist in the asymptotic
states of this model. The latter work has also shown, in accordance
with the expected confining behavior, that electrons arise in deep-inelastic-scattering
scenarios - i.e., short distances and high energies, and they behave
like free pointlike particles (agreeing with the {}``parton model'',
or the asymptotic freedom of $QCD$, where the confined quarks are
asymptotically free within the hadrons - at short ranges and high
energies). The existence of massive photons is in accordance with
the fact that in terms of statistical mechanics, confinement is a
gapped, disordered phase, while the ordered phases do not confine
\cite{Fradkin1978,Polyakov1987}.

A significant step in the study of confinement was done by Wilson,
who introduced lattice gauge theory for this purpose \cite{Wilson}.
Using loops in lattice gauge theory, which were later called {}``Wilson
Loops'', he showed that 3+1 dimensional compact $QED$ has a confining
phase (for strong coupling, $g^{2}\gg1$), which binds charges together
and forbids the existence of free electrons and positrons, as well
as a Coulomb phase (for weak coupling, $g^{2}\ll1$), as in continuous
$QED$. It was shown later (see subsection \ref{sub:The-Weak-Limit})
that 2+1 dimensional $cQED$ confines for all values of the coupling
constant \cite{Polyakov}: confinement in compact $QED$ was identified
as a topological effect, having to do with instantons and magnetic
monopoles. This result was supported by other works as well \cite{BanksMyersonKogut,DrellQuinnSvetitskyWeinstein,BenMenahem}.

Similarly to the {}``charge confinement'' in Abelian theories, confinement
also takes place in non-Abelian gauge theories, such as non-Abelian
Yang-Mills theories (as $QCD$) as well as the discrete theory of
$\mathbb{Z}_{N}$ presented earlier. It was argued \cite{Hooft1978}
that large-distance phenomena, involving the confinement of quarks
in $QCD$, is related with the group\textquoteright{}s center - $\mathbb{Z}_{3}$,
which makes the phase structure of $\mathbb{Z}_{N}$ interesting \cite{Horn1979,Elitzur1979,Bhanot1980}:
The theory confines in the strong coupling limit ($\lambda\rightarrow\infty$).
In 2+1 and 3+1 dimensions it is not the only phase. In 3 + 1 dimensions,
due to the self-duality of the $\mathbb{Z}_{N}$ theory, there is
a phase transition from electric confinement (strong coupling) to
magnetic confinement (weak coupling), for $N<N_{c}\quad(N_{c}\sim6)$.
For $N>N_{c}$ there is a third phase, with no confinement at all.
In 2+1 dimensions there is a phase transition to a non-confining phase
in the weak limit \cite{Fradkin1978,Horn1979}. However, for $N\rightarrow\infty$
the theory shows the phase transition at $g=0$ \cite{Bhanot1980},
in accordance with the single confining phase structure of $U(1)$
(the $N\rightarrow\infty$ limit of $\mathbb{Z}_{N}$).

Both in $U(1)$ and $SU(N)$, one can show, using perturbation theory,
the confinement of charges/quarks in the strong coupling limit, as
will be done in subsection \ref{sub:Strong-Coupling-Lattice}. The
confinement of static charges is manifested by {}``electric flux
tubes'' connecting them, which get thicker as the coupling constant
decreases. But, outside the strong limit, confinement (in cases it
holds) becomes nontrivial and requires nonperturbative techniques,
and for Yang-Mills theories it is still an open question: in fact,
the proof and calculation of the Mass Gap in Yang-Mills theories,
which is related to quark confinement, is one of Clay's institute
Millennium Prize problems. For example, in the cases confinement was
proven, it was mostly done for static charges - and not dynamical fermions.

Dynamical fermions shall enhance electric flux-tube breaking. If one
stretches a flux tube, at a certain point it would be energetically
favorable to create two pair of {}``new'' quarks, or charges, breaking
the a long flux tube into two smaller ones.

\subsubsection{\label{sub:Strong-Coupling-Lattice}Strong Coupling Lattice Gauge
Theories.}

An interesting limit of lattice Yang-Mills theories is the strong
coupling limit $g^{2}\gg1$, in which the Kogut-Susskind Hamiltonian
(\ref{eq:KS_Ham}) may be treated perturbatively: the magnetic, plaquette
part $H_{B}$ may be treated as a perturbation to the electric $H_{E}\equiv H_{0}$.
\footnote{It is worth to comment that strong coupling calculations were initially
performed in the Euclidean formalism. The first to do that was Wilson
\cite{Wilson}.
}

Consider two opposite static charges, placed on the vertices $\mathbf{n}$
and $\mathbf{n}+R\mathbf{\hat{x}}$ of a lattice. What is the ground
state of this static charge configuration in the strong coupling limit?
In this limit, products of local (links) electric field states form
a good basis, as they are eigenstates of $H_{E}$, and thus we obviously
must find the electric field configuration with minimal energy, which
satisfies the \emph{static }Gauss's law for these two charges.

The no-charge vacuum of this limit satisfies
\begin{equation}
\left|\Omega\left(g^{2}\gg1\right)\right\rangle =\underset{\mathbf{n},k}{\otimes}\left|0\right\rangle
\end{equation}
- on each link the state satisfies

\begin{equation}
\mathbf{L}^{2}\left|0\right\rangle =0
\end{equation}
- no electric field at all.

If we excite a link (using $U$) the zeroth order energy is increased
by a unit of $\mathbf{L}^{2}$:
\begin{equation}
\mathbf{L}^{2}U\left|0\right\rangle =\left[\mathbf{L}^{2},U\right]\left|0\right\rangle =\left(\underset{a}{\sum}T_{i}^{\left(r\right)}T_{i}^{\left(r\right)}\right)U\left|0\right\rangle
\end{equation}
where $T_{i}^{r}$ is the matrix representation of the group's generator
in representation $r$, and $C_{2}\left(r\right)\equiv\underset{i}{\sum}T_{i}T_{i}$
is just a C-number - the eigenvalue of the Casimir operator in this
representation.

Thus, for example, in the case of $cQED$, we get \cite{KogutLattice}
\begin{equation}
L^{2}U\left|0\right\rangle =U\left|0\right\rangle
\end{equation}
and in the case of $SU\left(2\right)$, for charges in the fundamental
representation $j=\frac{1}{2}$ \cite{Kogut1983},
\begin{equation}
\mathbf{L}^{2}U\left|0\right\rangle =\left(\frac{1}{4}\underset{a}{\sum}\sigma_{a}\sigma_{a}\right)U\left|0\right\rangle =\frac{3}{4}U\left|0\right\rangle
\end{equation}
or in any other representation $r$,
\begin{equation}
\mathbf{L}^{2}U^{r}\left|0\right\rangle =j_{r}\left(j_{r}+1\right)U^{r}\left|0\right\rangle
\end{equation}

This means that if we have $N$ links with a single excitation, the
energy of the state (in zeroth order) will be $\frac{Ng^{2}}{2}C_{2}\left(r\right)$.
In order to respect Gauss's law (and fulfill gauge invariance), the
eigenstates of $H_{E}$ in this charge configuration will contain
a flux line of $U\left|0\right\rangle $ states, ranging between the
charges. The minimal energy is obtained for the direct line, and thus
 the ground state is:
\begin{equation}
\left|R_{0,mm'}\right\rangle =\left(\underset{\mathbf{n}=\mathbf{n}}{\overset{\mathbf{n+}\left(R-1\right)\mathbf{\mathbf{\hat{x}}}}{\prod}}U_{\mathbf{n},x}\right)_{mm'}\left|\Omega\right\rangle
\end{equation}
\footnote{The indices in the state are another manifestation of the fact we
disregard the charge degrees of freedom in this context. If one introduces
the matter operators $\psi_{\mathbf{n}}$, the operator creating the
state out of the full vacuum would be $\underset{mm'}{\sum}\psi_{\mathbf{n},m}^{\dagger}\left(\underset{\mathbf{n}=\mathbf{n}}{\overset{\mathbf{n+}\left(R-1\right)\mathbf{\mathbf{\hat{x}}}}{\prod}}U_{\mathbf{n},x}\right)_{mm'}\psi_{\mathbf{\mathbf{n}+R\mathbf{\hat{x}}},m'}=\psi_{\mathbf{n}}^{\dagger}\underset{\mathbf{n}=\mathbf{n}}{\overset{\mathbf{n+}\left(R-1\right)\mathbf{\mathbf{\hat{x}}}}{\prod}}U_{\mathbf{n},x}\psi_{\mathbf{\mathbf{n}+R\mathbf{\hat{x}}}}$
which we easily identify as a gauge invariant string.}.
This flux-line is the so-called \emph{electric flux tube}, manifesting
confinement.

The zeroth order energy of this state is
\begin{equation}
E_{0}\left|R_{0,mm'}\right\rangle =H_{E}\left|R_{0,mm'}\right\rangle =\frac{g^{2}}{2}C_{2}\left(r\right)R\left|R_{0,mm'}\right\rangle
\end{equation}
and thus we get that the strong coupling limit confines, with a \emph{string
tension} (the ratio between the static potential and the distance)
of
\begin{equation}
\sigma_{strong}=\frac{g^{2}}{2}C_{2}\left(r\right)
\end{equation}

Such a flux-tube connecting the appropriate pair of them is nothing
but a \emph{meson}. One could also create a long flux tube connecting
static charges and watch it dynamically breaking into shorter, energy
favorable tubes, by dynamic pair creation. Also, as mentioned before,
outside the strong limit, but within the confining phase, the state
shall be modified such that the flux-tube is thicker (wider than just
a thin, single lattice line).

\subsubsection{\label{sub:The-Weak-Limit}The Weak Limit of $cQED$ .}

As $g$ decreases, perturbative corrections must be taken into account,
until perturbation theory breaks down and nonperturbative methods
of the weak regime must be utilized. As an example, we shall
comment about the weak limit of $cQED$.

In 3+1 dimensions, one can use continuum limit Wilson loop arguments
\cite{KogutLattice}, as the weak limit corresponds to the continuum
limit, and obtain the Coulomb potential.

In 1+1 dimensions, on the other hand, the theory confines for all
values of the coupling constant, as it is solvable and has no dependence
on the coupling constant $g$. This was shown (for the massless case)
in the continuum by Schwinger \cite{Schwinger1962I,Schwinger1962},
but similar results apply also in the lattice compact case in \cite{Banks1976}
(as well as in other, more recent, lattice studies, such as using
tensor network states \cite{Banuls2013,Banuls2013a,Kuhn2014,Buyens2014,Saito2014,Banuls2015,Pichler2015}).

In 2+1 dimensions interesting physics takes place, involving instantons
and topological effects. Due to magnetic monopoles, there is a mass
gap in the weak limit, corresponding to quark confinement. This was
shown by Polyakov \cite{Polyakov} using Euclidean partition functions,
by Drell, Quinn, Svetitsky and Weinstein, \cite{DrellQuinnSvetitskyWeinstein}
using variational Hamiltonian approach, and by Banks, Myerson and
Kogut \cite{BanksMyersonKogut}, using the Villain approximation.
The three approaches were compared by Ben-Menahem \cite{BenMenahem}
who also demonstrated that compactness is necessary for confinement
in lattice $QED$.

In some sense, the confinement mechanism of 2+1 dimensional $cQED$
is more related to the confinement of quarks in $QCD$, as both of
them are topological effects.

The discussion above applies for zero temperature. For a finite, nonzero
temperature, a Coulomb phase arises for weak values of the coupling
constant, with a phase transition to the Coulomb phase at some value
of $T,g$ \cite{Svetitsky,SvetitskyReview}.

\section{\label{sec:Ultracold-Atoms-in}Ultracold Atoms in Optical Lattices : A brief review }

\subsection{Generation of an Optical Potential}

Optical lattices are created by ``external'', classical lasers,
and virtual processes to an excited atomic state \cite{Jaksch1998,Jaksch2005,Bloch2008,Lewenstein2012}.

\begin{figure}
\begin{centering}
\includegraphics[scale=0.5]{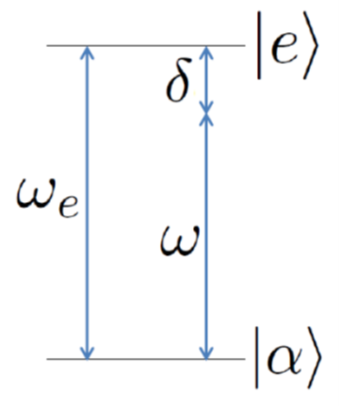}
\par\end{centering}

\protect\caption{The energy levels of the trapped atom. $\left|\alpha\right\rangle $
is the meta-stable level, for which the optical potential is effectively
obtained by eliminating the excited level $\left|e\right\rangle $.
The levels are separated by frequency $\omega_{e}$, and they are
coupled non-resonantly with a laser of frequency $\omega$: the detuning
is $\delta=\omega_{e}-\omega$.\label{fig:The-energy-levels}}
\end{figure}

Consider a two level atom (disregard the other levels), whose levels
are $\left|\alpha\right\rangle $ - some meta-stable level (in which
the atom is initially prepared), and $\left|e\right\rangle $ - an
excited one (see figure \ref{fig:The-energy-levels}). We may write
its Hamiltonian in the form
\begin{equation}
H_{atom}=\frac{\mathbf{p}^{2}}{2m}+\omega_{e}\left|e\right\rangle \left\langle e\right|
\end{equation}
The atoms are usually alkaline, therefore they have a single valence electron,
which simplifies their treatment by many aspects.

The atom experiences an external, classical laser field, of the form
\begin{equation}
\mathbf{E}\left(\mathbf{x},t\right)=E\left(\mathbf{x}\right)e^{-i\omega t}\hat{\mathbf{\epsilon}}
\end{equation}
Switching to a system rotating with the laser's frequency $\omega$,
and defining the detuning $\delta=\omega_{e}-\omega$ (see figure
\ref{fig:The-energy-levels}), we get
\begin{equation}
H_{atom}=\frac{\mathbf{p}^{2}}{2m}+\delta\left|e\right\rangle \left\langle e\right|
\end{equation}

The interaction between the laser and the atom is given in the dipole
approximation, which is valid if we assume that the electric field's
amplitude varies slowly compared to the atomic size (in space) and
$1/\omega$ (in time;
Note that here, for simplicity, the amplitude was not even considered
as having a time dependence, but one may generalize to a time-dependent
amplitude, of the form $E\left(\mathbf{x},t\right)$). This interaction has the form
\begin{equation}
H_{dipole}=-\mathbf{d}\cdot\mathbf{E}\left(\mathbf{x},t\right)
\end{equation}
where $\mathbf{d}$ is the dipole moment of the operator; After transforming
to the rotating frame, performing a rotating wave approximation (neglecting
the fast-rotating terms) and plugging $d_{ij}=\hat{\mathbf{\epsilon}}\cdot\left\langle i\right|\mathbf{d}\left|j\right\rangle $
, the dipole moment matrix elements (projected on the laser's polarization),
one obtains the interaction Hamiltonian
\begin{equation}
H_{dipole}=-d_{e\alpha}E\left(\mathbf{x}\right)\left|e\right\rangle \left\langle \alpha\right|+h.c.\equiv\frac{\Omega\left(\mathbf{x}\right)}{2}\left|e\right\rangle \left\langle \alpha\right|+h.c.
\end{equation}
where
\begin{equation}
\Omega\left(\mathbf{x}\right)=-2E\left(x\right)d_{e\alpha}
\end{equation}
is the Rabi frequency ($d_{e\alpha}=d_{\alpha e}$, as the dipole
operator is Hermitian).

We assume that the detuning $\delta$ is sufficiently large - the
coupling is non-resonant, and that transitions between the atomic
levels are practically impossible. However, virtual second-order transitions
to the excited level and back are possible, and thus
we get the effective Hamiltonian (for the $\left|\alpha\right\rangle $
subspace)
\begin{equation}
H=\frac{\mathbf{p}^{2}}{2m}+V_{op}\left(\mathbf{x}\right)
\end{equation}
where
\begin{equation}
V_{op}\left(\mathbf{x}\right)=-\frac{\left|\Omega\left(\mathbf{x}\right)\right|^{2}}{4\delta}
\end{equation}
is called the optical trapping potential, and it is an effective potential
experienced by the atom in the $\left|\alpha\right\rangle $ state.
In fact, this is merely the AC Stark effect. Note that we have actually
``traced out'' the internal levels, due to the energy restriction
(large detuning) and what we now have is a configuration space Hamiltonian.
Choosing the effective field properly, one could generate an optical
lattice: a periodic trapping potential, allowing the trapped atoms
to fill its minima. Since $V_{op}\left(\mathbf{x}\right)\propto\left|E\left(\mathbf{x}\right)\right|^{2}$,
a standing wave of the form $E\left(x\right)=E_{0}\cos\left(kx\right)$
would result, for example, in an optical potential of the form $V_{op}\left(x\right)=V_{0}\cos^{2}\left(kx\right)$;
One could add lasers in more than one direction to obtain a two- and
three-dimensional optical lattices of various geometries.

Next we shall consider an optical lattice with many atoms. This shall
be done in the context of second quantization, for which we are interested
in the single-particle eigenfunctions of the trapping Hamiltonian.

Due to the lattice symmetry, the single particle wavefunctions are
merely Bloch waves. These states are not localized, therefore if one
wishes to work in the number basis of local minima (Fock basis), a
transformation to local single particle wavefunctions is required.
This is fulfilled by the orthonormal basis of Wannier functions, $w_{n}\left(\mathbf{x}-\mathbf{x}_{0}\right)$,
which are the discrete Fourier transform of the Bloch wavefunctions
$u_{\mathbf{q}}^{\left(n\right)}\left(\mathbf{x}\right)$ over the
lattice \cite{Jaksch2005,Bloch2008,Lewenstein2012}:
\begin{equation}
w_{n}\left(\mathbf{x}-\mathbf{x}_{0}\right)\propto\underset{\mathbf{q}}{\sum}e^{-i\mathbf{q\cdot\mathbf{x}_{0}}}u_{\mathbf{q}}^{\left(n\right)}\left(\mathbf{x}\right)
\end{equation}
 These states are labeled by $n$, which is the energy band, and are
each spatially centered around a different local minimum $\mathbf{x}_{0}$
of the optical potential. These are the desired local site (minimum)
wavefunctions.

Working in low enough temperatures, it is reasonable to include only
the lowest energy band, $w_{0}\left(\mathbf{x}-\mathbf{x}_{0}\right)$.

One can expand the second-quantization ``wavefunction'' of the condensates
in terms of the localized Wannier functions $\phi_{\mathbf{n}}\left(\mathbf{x}\right)\equiv w_{0}\left(\mathbf{x}-\mathbf{x}_{\mathbf{n}}\right)$
and single particle annihilation operators $a_{\mathbf{n}}$ (bosonic
or fermionic, depending on the statistics of the atoms), both defined
at the site (minimum of the potential) $\mathbf{n}$, as
\begin{equation}
\Phi\left(\mathbf{x}\right)=\underset{\mathbf{n}}{\sum}a_{\mathbf{n}}\phi_{\mathbf{n}}\left(\mathbf{x}\right)\label{eq:Wanexp}
\end{equation}
and construct from which a usual second quantization Hamiltonian density:
\begin{equation}
\mathcal{H}=\Phi^{\dagger}\left(\mathbf{x}\right)\left(-\frac{\nabla^{2}}{2m}+V_{op}\left(\mathbf{x}\right)+V_{T}\left(\mathbf{x}\right)\right)\Phi\left(\mathbf{x}\right)+\int d^{3}x'\int d^{3}x\Phi^{\dagger}\left(\mathbf{x'}\right)\Phi^{\dagger}\left(\mathbf{x}\right)V\left(\mathbf{x-\mathbf{x'}}\right)\Phi\left(\mathbf{x}\right)\Phi\left(\mathbf{x'}\right)\label{eq:Hden}
\end{equation}
where $V_{T}\left(\mathbf{x}\right)$ is an external trapping potential,
varying slowly compared to the optical lattice potential $V_{op}\left(\mathbf{x}\right)$,
and $V\left(\mathbf{x-\mathbf{x'}}\right)$ is some two-body interaction,
which will be specified later.

\subsection{Multi-species Lattices and Spinor Condensates}

It is also possible to utilize more atomic levels and transitions
and create a ``species-dependent lattice'': i.e., with different
minima for different atomic species, which may be, for example, different
atomic hyperfine levels \cite{Ketterle2001,Pethick2002,Jaksch2005,Lewenstein2007,Lewenstein2012}.

The atomic hyperfine structure is given by an $SU(2)$ Hilbert space,
characterized by two quantum numbers $F,m_{F}$. Just like ``regular''
angular momentum,
\begin{equation}
\mathbf{F}^{2}\left|F,m_{F}\right\rangle =F\left(F+1\right)\left|F,m_{F}\right\rangle
\end{equation}
and
\begin{equation}
F_{z}\left|F,m_{F}\right\rangle =m_{F}\left|F,m_{F}\right\rangle
\end{equation}

$\mathbf{F}$ is defined as
\begin{equation}
\mathbf{F=I+L+S}
\end{equation}
 where $\mathbf{I}$ is the nuclear spin, and $\mathbf{L,S}$ stand
for the valence electron's orbital angular momentum and spin respectively
(recall that we are dealing with alkaline atoms, with a single valence
electron), and thus the Lie algebra
\begin{equation}
\left[F_{i},F_{j}\right]=i\epsilon_{ijk}F_{k}
\end{equation}
 is satisfied. Each species $\left|\alpha\right\rangle $ can then
experience a different optical potential $V_{op}^{\alpha}\left(\mathbf{x}\right)$,
and these potentials form a \emph{superlattice}.

Such lattices / condensates are sometimes referred to as spinor lattices
/ condensates\emph{. }

As an example, consider $^{87}\mathrm{Rb}$ and $^{23}\mathrm{Na}$
have a nuclear spin $I=\frac{3}{2}$. Thus, in the S-wave hyperfine
manifold (which is relevant for ultracold atoms, as explained in subsection
\ref{sub:Scattering-of-Ultracold}), $F=1$ or $2$.

We denote the internal levels by Greek indices, and the second-quantization
expansion generalizes to
\begin{equation}
\Phi_{\alpha}\left(\mathbf{x}\right)=\underset{\mathbf{n}}{\sum}a_{\mathbf{n},\alpha}\phi_{\mathbf{n},\alpha}\left(\mathbf{x}\right)
\end{equation}
for each species, where the Wannier functions may be species dependent
or independent - depending on the lattice structure and symmetry and
the choice of these levels.

The Hamiltonian density is now
\begin{equation}
\begin{aligned}\mathcal{H} & = & \underset{\alpha,\beta}{\sum}\Phi_{\alpha}^{\dagger}\left(\mathbf{x}\right)\left(\delta^{\alpha\beta}\left(-\frac{\nabla^{2}}{2m}+V_{op}^{\alpha}\left(\mathbf{x}\right)+V_{T}\left(\mathbf{x}\right)\right)+\Omega^{\alpha\beta}\left(\mathbf{x}\right)\right)\Phi_{\beta}\left(\mathbf{x}\right)\\
 & + & \underset{\alpha,\beta,\gamma,\delta}{\sum}\int d^{3}x'\Phi_{\alpha}^{\dagger}\left(\mathbf{x'}\right)\Phi_{\beta}^{\dagger}\left(\mathbf{x}\right)V_{\alpha\beta\gamma\delta}\left(\mathbf{x-\mathbf{x'}}\right)\Phi_{\gamma}\left(\mathbf{x}\right)\Phi_{\delta}\left(\mathbf{x'}\right)
\end{aligned}
\end{equation}
where $\Omega^{\alpha\beta}\left(\mathbf{x}\right)$ corresponds to
a Rabi coupling of atomic levels by more lasers.

\subsection{\label{sub:Scattering-of-Ultracold}Scattering of Ultracold Atoms}

How cold is ultracold? The ultracold regime is defined as the one
in which the lowest partial wave governs the scattering. It is a dilute
and cold regime.

We are dealing with the so called \emph{weak interaction regime},
involving the ground state of a gas of either bosonic or fermionic
atoms. For ideal bosons, we assume that they all occupy the lowest
energy band (Bose Einstein Condensate). A phase transition from Bose-Einstein
condensate takes place in a finite temperature (in three dimensions),
when the thermal de-Broglie wavelength $\lambda_{T}=\sqrt{\frac{2\pi}{k_{B}T}}$
(in $\hbar=1$ units) approaches $n^{-1/3}$ - the average inter-particle
distance ($n$ is the number density). For fermions, the states are
full up to the Fermi energy, $\epsilon_{F}$: for temperatures satisfying
$k_{B}T\ll\epsilon_{F}$ the gas is considered degenerate \cite{Bloch2008}.

When the energy is low enough - i.e., the de-Broglie wavelength of
the atoms is much larger than the effective range of the potential
\cite{Lewenstein2012}, the interaction potential in (\ref{eq:Hden})
is accurately described by the pseudopotential:
\begin{equation}
V\left(\mathbf{x-\mathbf{x'}}\right)=\frac{2\pi a}{m}\delta^{\left(3\right)}\left(\mathbf{x-\mathbf{x'}}\right)\equiv\frac{g}{2}\delta^{\left(3\right)}\left(\mathbf{x-\mathbf{x'}}\right)
\end{equation}
For bosons the scattering is governed by the S-wave channel. For identical
fermions, on the other hand, due to Pauli's exclusion principle, there
is no S-wave scattering \cite{Bloch2008}; This is not a problem for
the scattering of two different fermions (non-identical).

The density of the atoms, $n$, is typically $10^{12}-10^{15}\mathrm{cm}^{-3}$.
This gives a typical interparticle distance $n^{-1/3}$ of the order
of $0.1-10\mathrm{\mu m}$, and the scattering length is in the range
of few nanometers \cite{Bloch2008}. The scattering length of $^{87}\mathrm{Rb}$,
for example, has on order of magnitude of 100 times the Bohr radius,
varying slightly according to the different internal states \cite{Pethick2002}.
The gas is sufficiently dilute for $n^{1/3}a\ll1$, in case of bosonic
repulsive interactions ($a>0$), as $\sqrt{na^{3}}$ is the small
parameter of the Bogolyubov theory, describing weakly interacting
Bose gases \cite{Bloch2008}.

We thus obtain a simplified version for the single-species case Hamiltonian
density:
\begin{equation}
\mathcal{H}=\Phi^{\dagger}\left(\mathbf{x}\right)\left(-\frac{\nabla^{2}}{2m}+V_{op}\left(\mathbf{x}\right)+V_{T}\left(\mathbf{x}\right)\right)\Phi\left(\mathbf{x}\right)+\frac{g}{2}\Phi^{\dagger}\left(\mathbf{x}\right)\Phi^{\dagger}\left(\mathbf{x}\right)\Phi\left(\mathbf{x}\right)\Phi\left(\mathbf{x}\right)
\end{equation}
after plugging in the Wannier functions and creation/annihilation
operators (\ref{eq:Wanexp}) one may integrate over the Wannier functions.
Using the overlap integrals
\begin{equation}
\epsilon_{n}=\int d^{3}x\phi_{n}^{*}\left(\mathbf{x}\right)\left(-\frac{\nabla^{2}}{2m}+V_{op}\left(\mathbf{x}\right)+V_{T}\left(\mathbf{x}\right)\right)\phi_{n}\left(\mathbf{x}\right)\label{eq:epsilon}
\end{equation}
\begin{equation}
J_{mn}=\int d^{3}x\phi_{m}^{*}\left(\mathbf{x}\right)\left(-\frac{\nabla^{2}}{2m}+V_{op}\left(\mathbf{x}\right)+V_{T}\left(\mathbf{x}\right)\right)\phi_{n}\left(\mathbf{x}\right)\label{eq:J}
\end{equation}
\begin{equation}
U_{mnkl}=g\int d^{3}x\phi_{m}^{*}\left(\mathbf{x}\right)\phi_{n}^{*}\left(\mathbf{x}\right)\phi_{k}\left(\mathbf{x}\right)\phi_{l}\left(\mathbf{x}\right)\label{eq:U}
\end{equation}
 where $m,n,k,l$ label the lattice sites. The most general Hamiltonian,
taking into account all possible overlap integrals, is

\begin{equation}
H=\underset{m,n}{\sum}J_{m,n}a_{m}^{\dagger}a_{n}+\underset{m,n,k,l}{\sum}U_{m,n,k,l}a_{m}^{\dagger}a_{n}^{\dagger}a_{k}a_{l}
\end{equation}

\subsection{\label{sub:Multiple-Species-Scattering}Multiple Species Scattering}

In case of multiple species, the most general form of the Hamiltonian
is
\begin{eqnarray}
H & = & \underset{m,n,\alpha,\beta}{\sum}J_{m,n}^{\alpha,\beta}a_{m,\alpha}^{\dagger}a_{n,\beta}+\underset{m,n,k,l,\alpha,\beta,\gamma,\delta}{\sum}U_{m,n,k,l}^{\alpha,\beta,\gamma,\delta}a_{m,\alpha}^{\dagger}a_{n,\beta}^{\dagger}a_{k,\gamma}a_{l,\delta}
\end{eqnarray}
however, if we remember that the species are different hyperfine states,
and take into account the hyperfine rotational symmetry (conservation
of the hyperfine angular momentum in atomic collisions) we can simplify
this Hamiltonian further, as we shall do next \cite{Ketterle2001,Pethick2002}.

Suppose we wish to consider the scattering of two atoms with distinct
$F,m_{F}$ values: for example, let us scatter an $F_{1}$ atom on
an $F_{2}$ atom. As a result, we shall get a pair of two atoms with
their own $F,m_{F}$ values. However, taking into account the rules
of addition of angular momenta, one should be aware that the scattering
may take place ``through'' an ``intermediate state'' with several
values of total angular momentum $F_{T}$. Thus, the effective scattering
potential will take the form
\begin{equation}
V_{\alpha,\beta,\gamma,\delta}\left(\mathbf{x-\mathbf{x'}}\right)=\frac{2\pi}{m}\delta^{\left(3\right)}\left(\mathbf{x-\mathbf{x'}}\right)\underset{F_{T}}{\sum}a_{F_{T}}\left(P_{F_{T}}\right)_{\alpha,\beta,\gamma,\delta}
\end{equation}
where we sum over the possible values of total angular momentum $F_{T}$,
and consider, for each value, its corresponding scattering length
$a_{F_{T}}$. $P_{F_{T}}$ is the projection operator onto the subspace
of total angular momentum $F_{T}$. In case of different atomic masses,
the appropriate reduced masses should replace the mass $m$.

The projection operators can be represented using functions of $\mathbf{F_{1}}\cdot\mathbf{F_{2}}$.
Physically, that may be understood as a consequence of rotational
invariance: terms which depend on the $F$ values of two atoms and
conserve the hyperfine angular momentum must be constructed out of
powers of these terms. Mathematically,
\begin{equation}
\mathbf{F_{1}}\cdot\mathbf{F_{2}}=\frac{1}{2}\left(\mathbf{F}_{T}^{2}-\mathbf{F}_{1}^{2}-\mathbf{F}_{2}^{2}\right)
\end{equation}
and as $\mathbf{F}_{1}^{2}$, $\mathbf{F}_{2}^{2}$ are practically
constants for our purposes, a function of $\mathbf{F_{1}}\cdot\mathbf{F_{2}}$
is in fact a function of $\mathbf{F}_{T}^{2}$, and thus one can construct
the projection operators out of this scalar product. If there are
$n$ possible values of total hyperfine angular momentum, the projection
operators are polynomials of the form
\begin{equation}
P_{F_{T}}=\underset{k=0}{\overset{n-1}{\sum}}G_{F_{T},k}\left(\mathbf{F_{1}}\cdot\mathbf{F_{2}}\right)^{k}
\end{equation}
and eventually obtain
\begin{equation}
V_{\alpha,\beta,\gamma,\delta}\left(\mathbf{x-\mathbf{x'}}\right)=\delta^{\left(3\right)}\left(\mathbf{x-\mathbf{x'}}\right)\frac{g_{k}}{2}\left(\left(\mathbf{F_{1}}\cdot\mathbf{F_{2}}\right)^{k}\right)_{\alpha,\beta,\gamma,\delta}
\end{equation}
where the $\left\{ g_{k}\right\} $ coefficients depend simply on
the scattering lengths $\left\{ a_{F_{T}}\right\} $ and the coefficients
in the projection operators' expansion, $\left\{ G_{F_{T},k}\right\} $.
involved. The scattering lengths are tunable
using Feshbach resonances, either magnetic or optical \cite{Feshbach1958,Fedichev1996,Bohn1997,Inouye1998,Cornish2000,Fatemi2000,Pethick2002,Bloch2008,Chin2010,Blatt2011}.

We can also express the projection operators, and thus the scattering
potential, in terms of the Clebsch-Gordan coefficients:
\begin{equation}
\begin{aligned}
&V_{\alpha,\beta,\gamma,\delta}\left(\mathbf{x-\mathbf{x'}}\right)  =  \delta^{\left(3\right)}\left(\mathbf{x-\mathbf{x'}}\right)\underset{F_{T}}{\sum}C_{F_{T}}\left\langle F_{1},m_{F1}=\alpha;F_{2},m_{F2}=\beta\right|\left.F_{T},M_{F}\right\rangle \times \\
 &  \left\langle F_{T},M_{F}\right|\left.F_{1},m_{F1}=\gamma;F_{2},m_{F2}=\delta\right\rangle \nonumber
 \end{aligned}
\end{equation}
which explicitly manifests, as a consequence of hyperfine angular
momentum conservation, that $\alpha+\beta=\gamma+\delta$ , and the
$\left\{ C_{F_{T}}\right\} $ coefficients depend on the scattering
lengths.

As an example \cite{Ketterle2001,Pethick2002}, one may consider a
spinor condensate of $F=1$. The possible values of total angular
momentum in collisions are naively $F_{T}=0,1,2$, but since the atoms
are bosons, and the states must be symmetric, we remain with $F_{T}=0,2$.
Thus the scattering lengths we need are $a_{0}$ and $a_{2}$, and
the highest power of $\mathbf{F_{1}}\cdot\mathbf{F_{2}}$ we need
is 1. The projection operators are
\begin{equation}
P_{0}=\frac{1}{3}\left(1-\mathbf{F_{1}}\cdot\mathbf{F_{2}}\right)
\end{equation}
and
\begin{equation}
P_{2}=\frac{1}{3}\left(\mathbf{F_{1}}\cdot\mathbf{F_{2}}+2\right)
\end{equation}

One can define
\begin{equation}
g_{0}=\frac{4\pi}{3m}\left(2a_{2}+a_{0}\right)
\end{equation}
and
\begin{equation}
g_{2}=\frac{4\pi}{3m}\left(a_{2}-a_{0}\right)
\end{equation}
and obtain the scattering part of the Hamiltonian density
\begin{equation}
\frac{g_{0}}{2}\underset{\alpha,\beta}{\sum}\Phi_{\alpha}^{\dagger}\left(\mathbf{x}\right)\Phi_{\beta}^{\dagger}\left(\mathbf{x}\right)\Phi_{\beta}\left(\mathbf{x}\right)\Phi_{\alpha}\left(\mathbf{x}\right)+\frac{g_{2}}{2}\underset{\alpha,\beta,\gamma,\delta}{\sum}\left(\Phi_{\alpha}^{\dagger}\left(\mathbf{x}\right)\mathbf{F}_{\alpha\gamma}\Phi_{\gamma}\left(\mathbf{x}\right)\right)\cdot\left(\Phi_{\beta}^{\dagger}\left(\mathbf{x}\right)\mathbf{F}_{\beta\delta}\Phi_{\delta}\left(\mathbf{x}\right)\right)
\end{equation}
where $\mathbf{F}$ are the $F=1$ representation matrices of $SU(2)$.

\subsection{Summary of the Available Interactions and Possibilities}
To conclude, in optical lattices one could use either bosonic or fermionic
ultracold atoms, of species $\left\{ \alpha\right\} $, with single
particle local wavefunctions at sites $n$ , $\phi_{n,\alpha}\left(\mathbf{x}\right)$
and $\psi_{n,\alpha}\left(\mathbf{x}\right)$ for bosons and fermions
respectively, and annihilation operators $a_{n,\alpha}$ and $c_{n,\alpha}$
for bosons and fermions respectively, satisfying the canonical commutation
relations
\begin{equation}
\left[a_{n,\alpha},a_{m,\beta}^{\dagger}\right]=\delta_{nm}\delta_{\alpha\beta}
\end{equation}
\begin{equation}
\left\{ c_{n,\alpha},c_{m,\beta}^{\dagger}\right\} =\delta_{nm}\delta_{\alpha\beta}
\end{equation}
\begin{equation}
\left[a_{n,\alpha},a_{m,\beta}\right]=\left\{ c_{n,\alpha},c_{m,\beta}\right\} =\left[a_{n,\alpha},c_{m,\beta}\right]=0
\end{equation}

Out of these operators and wavefunctions, the second-quantization
field operators are constructed,
\begin{equation}
\Phi_{\alpha}\left(\mathbf{x}\right)=\underset{n}{\sum}a_{n,\alpha}\phi_{n,\alpha}\left(\mathbf{x}\right)
\end{equation}
\begin{equation}
\Psi_{\alpha}\left(\mathbf{x}\right)=\underset{n}{\sum}c_{n,\alpha}\psi_{n,\alpha}\left(\mathbf{x}\right)
\end{equation}
and they are the ingredients of the atomic Hamiltonian, consisting
of three main parts:

\begin{enumerate}
\item The single-particle terms,
\begin{equation}
H_{0}=\underset{\alpha}{\sum}\int d^{3}\mathbf{x}\left(\Psi_{\alpha}^{\dagger}\left(\mathbf{x}\right)H_{0,f}\Psi_{\alpha}\left(\mathbf{x}\right)+\Phi_{\alpha}^{\dagger}\left(\mathbf{x}\right)H_{0,b}\Phi_{\alpha}\left(\mathbf{x}\right)\right)
\end{equation}
where $H_{0,f},H_{0,b}$ are the fermionic and bosonic single-particle
Hamiltonians, including the kinetic energy and the optical and trapping
potentials. They depend on (or, from another perspective, affect)
the shape of the lattice (through the optical potential) and thus
the single-particle wavefunctions, $\phi_{n,\alpha}\left(\mathbf{x}\right)$
and $\psi_{n,\alpha}\left(\mathbf{x}\right)$ .
\item The scattering terms,
\begin{eqnarray}
H_{sc} & = & \frac{1}{2}\underset{\alpha,\beta,\gamma,\delta}{\sum}g_{\alpha\beta\gamma\delta}^{FF}\int d^{3}\mathbf{x}\Psi_{\alpha}^{\dagger}\left(\mathbf{x}\right)\Psi_{\beta}^{\dagger}\left(\mathbf{x}\right)\Psi_{\gamma}\left(\mathbf{x}\right)\Psi_{\delta}\left(\mathbf{x}\right)\label{eq:Hsc}\\
 & + & \frac{1}{2}\underset{\alpha,\beta,\gamma,\delta}{\sum}g_{\alpha\beta\gamma\delta}^{BB}\int d^{3}\mathbf{x}\phi_{\alpha}^{\dagger}\left(\mathbf{x}\right)\phi_{\beta}^{\dagger}\left(\mathbf{x}\right)\phi_{\gamma}\left(\mathbf{x}\right)\phi_{\delta}\left(\mathbf{x}\right)\nonumber \\
 & + & \frac{1}{2}\underset{\alpha,\beta,\gamma,\delta}{\sum}g_{\alpha\beta\gamma\delta}^{BF}\int d^{3}\mathbf{x}\Psi_{\alpha}^{\dagger}\left(\mathbf{x}\right)\phi_{\beta}^{\dagger}\left(\mathbf{x}\right)\Psi_{\gamma}\left(\mathbf{x}\right)\phi_{\delta}\left(\mathbf{x}\right)\nonumber
\end{eqnarray}
which govern the two-body S-wave scattering (atomic collisions), and
depend on the scattering coefficients $g$. They depend on the scattering
lengths, and of course on the nature of collisions - fermion-fermion
(FF), boson-boson (BB) or boson-fermion (BF), and the atomic species
involved. These coefficients are not independent of each other, thanks
to the conservation of hyperfine angular momentum, and are tunable
using Feshbach resonances.
\item Rabi (laser) terms,
\begin{equation}
\begin{aligned}H_{R} & = & \underset{\alpha,\beta}{\sum}\Omega_{\alpha\beta}^{F}\int d^{3}\mathbf{x}\Psi_{\alpha}^{\dagger}\left(\mathbf{x}\right)\Psi_{\beta}\left(\mathbf{x}\right)\\
 & + & \underset{\alpha,\beta}{\sum}\Omega_{\alpha\beta}^{B}\int d^{3}\mathbf{x}\phi_{\alpha}^{\dagger}\left(\mathbf{x}\right)\phi_{\beta}\left(\mathbf{x}\right)
\end{aligned}
\label{eq:Rabi}
\end{equation}
which are governed by external lasers and may induce desired hopping
processes.
\end{enumerate}
One could also include transitions to and from molecular states, but
these are not of relevance for this report.

Current experimental technologies allow creating complex and controllable
optical lattices. The range of parameters is wide, as exemplified
by the typical numbers presented above: the ratio $V_{0}/E_{R}$ is
tunable. Optical lattices may be created in various geometries and
dimensionalities, using holographic techniques, for example. The lattices
may be very large (for example, 150000 lattice sites in \cite{Greiner2002})
and contain many atoms. The atoms may be distributed across the lattice
in various ways: lattice sites may occupy few atoms, even single ones
\cite{Greiner2002,Bakr2009,Weitenberg2011,Endres2012,Parsons2015}, or much more
(for example, 80 in \cite{Wurtz2009}, using tube-shaped sites).

Common measurements are the so-called ``time-of-flight'' measurements,
in which the lattice is switched off (suddenly or adiabatically) and
the atoms are released, in a way that their spatial density is proportional
to their momentum distribution, which is measurable using standard
imaging methods \cite{Bloch2008}. Contemporary experiments also demonstrate
with great success single-addressability of lattice sites and atoms
\cite{Wurtz2009,Bakr2009,Weitenberg2011,Parsons2015}, which opens the way for
various possibilities, both in measurements and initial state preparation.

Further detailed information on the topic may be found in \cite{Lewenstein2012}.

\section{A framework for Quantum Simulation of lattice gauge theories with ultracold atoms in optical lattices}
We shall review several methods for quantum
simulation - i.e., physical realization using ultracold atoms in optical
lattices, of three different lattice gauge theories: compact $QED$,
$SU(N)$ and $\mathbb{Z}_{N}$.

In order to simulate a gauge theory, three requirements must be fulfilled
\cite{AngMom}:
\begin{enumerate}
\item \emph{The theory must contain both fermions and bosons. }The fermions
will represent matter fields and the bosons - gauge fields. This makes
ultracold atoms natural candidates, as bosonic and fermionic atoms
are practically free resources in these systems. Ultracold atoms in optical lattices
are currently highly controllable systems.
Of course, the nature of lattice fermions should be chosen carefully, based on both
the experimental considerations and the desired theoretical approach for lattice fermions,
as discussed above.
\item \emph{The theory must be Lorentz invariant} - i.e., to have a causal
structure, as a field theory. This may seem problematic, but one can
simulate the lattice gauge theories described above,
that obtain this symmetry in the continuum limit. This goes along
quite well with the structure of the simulating systems, which are
optical \emph{lattices}.
\item \emph{The theory must have local gauge invariance}, which is the symmetry
responsible for gauge-matter interactions. This seems the most problematic
demand, and the biggest challenge, as this does not seem to be a fundamental
symmetry in such atomic systems, and has been one of the most important
research steps.
\end{enumerate}

Thus, after considering these three requirements and recognizing that
ultracold atoms in optical lattices should enable their fulfillment,
such systems have been chosen as the simulating systems - building
blocks for optical realizations of lattice gauge theories, opening
the way to quantum simulations of lattice gauge theories.
The main idea is to obtain the gauge invariant interactions which are not fundamental for ultracold atoms.
We present two methods for that - one in which gauge invariance is imposed as a constraint - the effective method,
in which the gauge invariance is only an emerging, low-energy sector symmetry,
and the other one, in which gauge invariance is mapped into a fundamental symmetry of the atoms - conservation
of hyperfine angular momentum in atomic collisions.

\begin{figure}[t]
\begin{centering}
\includegraphics[scale=0.6]{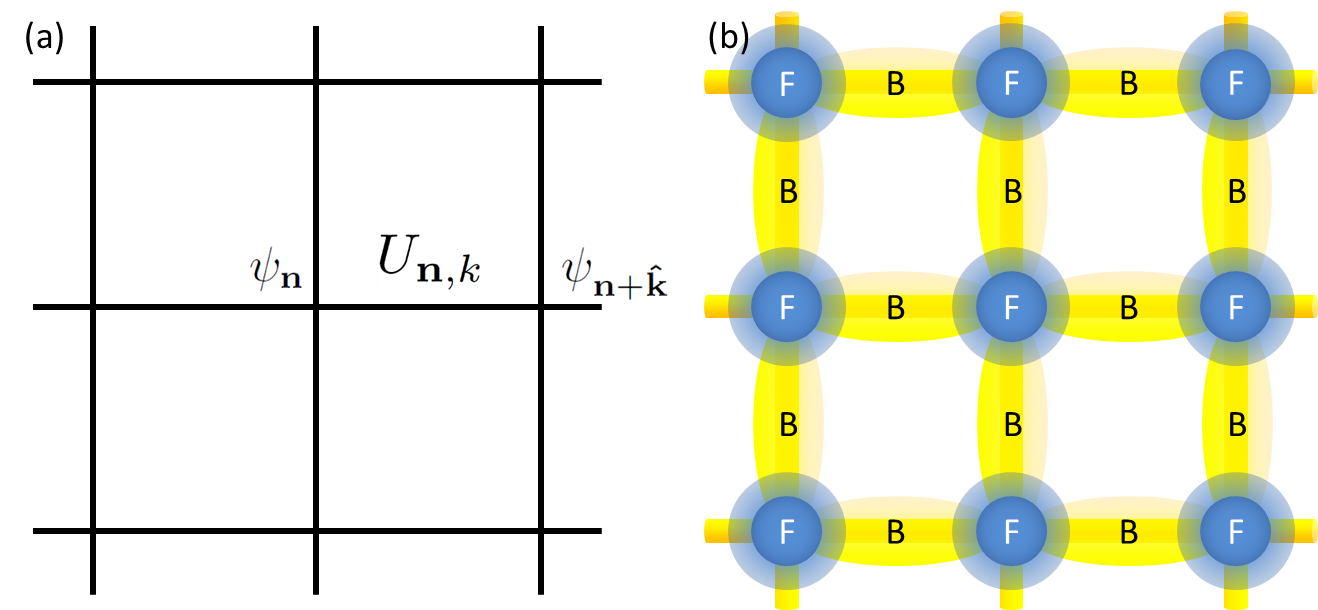}
\par\end{centering}
\caption{ (a) The lattice structure. The gauge degrees of freedom occupy the links, whereas the fermionic matter - the vertices.
This structure is kept in the atomic quantum simulator - whose schematic plot is presented in (b): the circles and
ellipses represent the minima of the optical potentials - Bosonic
(B, yellow) on the links, and Fermionic (F, blue) on the vertices.}
\label{optlat}
\end{figure}

In all the simulation approaches presented below, regardless of the
specific implementation and representation of the gauge degrees of freedom,
they are represented by bosons populating the links of an optical lattice.
Fermions, in proposals involving dynamical matter, are always located in minima coinciding
with the vertices (see figure \ref{optlat}b).

\section{Effective gauge invariance emerging at low energies}

\begin{figure}[t]
\begin{centering}
\includegraphics[scale=0.7]{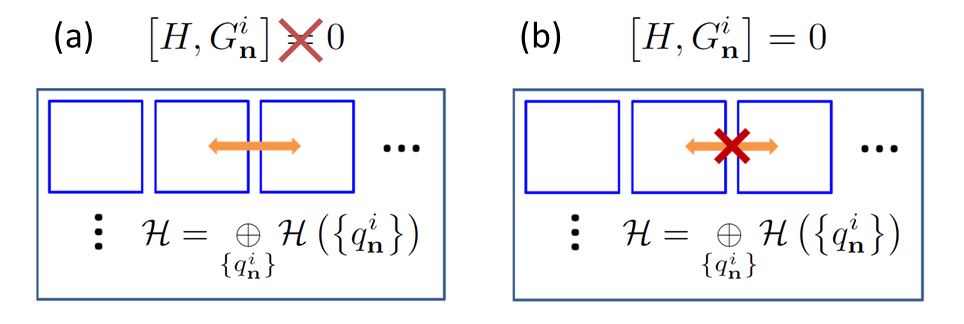}
\par\end{centering}

\caption{Effective gauge invariance. The original atomic Hamiltonian (a) mixes
among the Hilbert space sectors of different static charge configurations,
as it is not gauge invariant: the generators of gauge transformations
(\ref{eq:Gaugegen}) do not commute with the Hamiltonian. Using the
Gauss's law constraint (\ref{eq:HG}), one effectively eliminates
the transitions among these sectors, resulting in the desired gauge
invariant interactions, and the restoration of gauge invariance (\ref{eq:gausscomm})
by constructing an effective Hamiltonian to such a sector of interest.\label{fig:Effective-gauge-invariance.}}
\end{figure}

First we shall describe a class of models, in which the gauge symmetry is not fundamental,
but rather arises in a low energy sector, as an emerging, effective symmetry.
In this method, the atomic Hamiltonian is not gauge
invariant, but rather contains a large penalty term, demanding, as a
low energy constraint, that Gauss's law will be satisfied. The constraint
Hamiltonian takes the form
\begin{equation}
H_{G}=\lambda\underset{\mathbf{n}}{\sum}G_{\mathbf{n}}^{2}\label{eq:HG}
\end{equation}
where $\lambda$ is the largest energy scale in the system. This automatically
divides the physical Hilbert space into sectors, corresponding to
different eigenvalues of $G_{\mathbf{n}}$ (\ref{eq:Gaugegen}) -
different static charge configurations, but the Hamiltonian interactions
involve transitions among them. These are practically impossible,
due to energy conservation. Thus, an effective Hamiltonian \cite{Soliverez1969,CTBook} for the
ground sector has been introduced in any of these proposals, giving rise to a gauge-invariant
effective theory, which includes the desired interaction (see figure
\ref{fig:Effective-gauge-invariance.}).

\subsection{cQED in $2+1$ dimensions - link interactions}

It is shown in \cite{Zohar2011}, as a proof of principle, that a quantum
simulation of the Kogut-Susskind Hamiltonian for $cQED$ is, indeed,
possible, and this was the first work arguing that. This first proposal discussed
only the pure-gauge theory, suggesting a method to observe confinement
of static charges, by probing electric flux tubes.

In this work, every link of a two-dimensional spatial lattice (labeled
by the vertex from each it emanates $\left(m,n\right)$ and its direction
$k$) is occupied by a Bose-Einstein condensate, with a mean number
of atoms $N_{0}$ (uniform all over the lattice). We expand the local
number operators $N_{m,n}^{k}$,
\begin{equation}
N_{m,n}^{k}=N_{0}+\delta{}_{m,n}^{k}
\end{equation}
where $N_{0}$ is a C-number and $\delta{}_{m,n}^{k}$ - an operator,
measuring the excess of the atomic population over (or below) $N_{0}$
, which can take negative values as well - and represents the electric
field. These operators are conjugate to the local condensates' phases,
$\theta{}_{m,n}^{k}$, playing the role of the compact vector potential.

In a limit where the expectation value, as well as the uncertainty
of $\delta{}_{m,n}^{k}$ are much smaller than $N_{0}$, a quantum-rotor
approximation may be introduced, allowing the expansion of local annihilation
(or creation) operators as
\begin{equation}
a\approx\sqrt{N_{0}}e^{i\theta}
\end{equation}

\begin{figure}[t]
\begin{centering}
\includegraphics[scale=0.7]{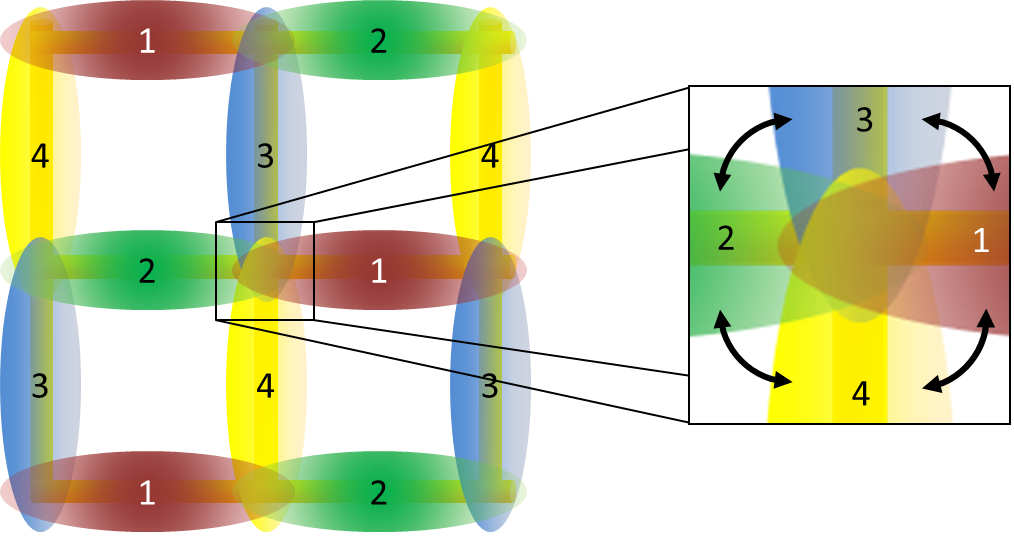}\caption{The lattice structure of \cite{Zohar2011}. Bosons of four different species (1-4) occupy the links
such that neighboring links are occupied by different species, and thus nearest-neighbor hopping is avoided. Instead, diagonal hopping processes
are induced by external lasers (denoted by black arrows on the right): $1\leftrightarrow3,3\leftrightarrow2,2\leftrightarrow4,4\leftrightarrow1$.
Scattering processes take place among the four species meeting in (and overlapping on) a vertex, and must be tailored properly.}\label{BECfig}

\par\end{centering}

\end{figure}

\begin{figure}[t]
\begin{centering}
\includegraphics[scale=0.9]{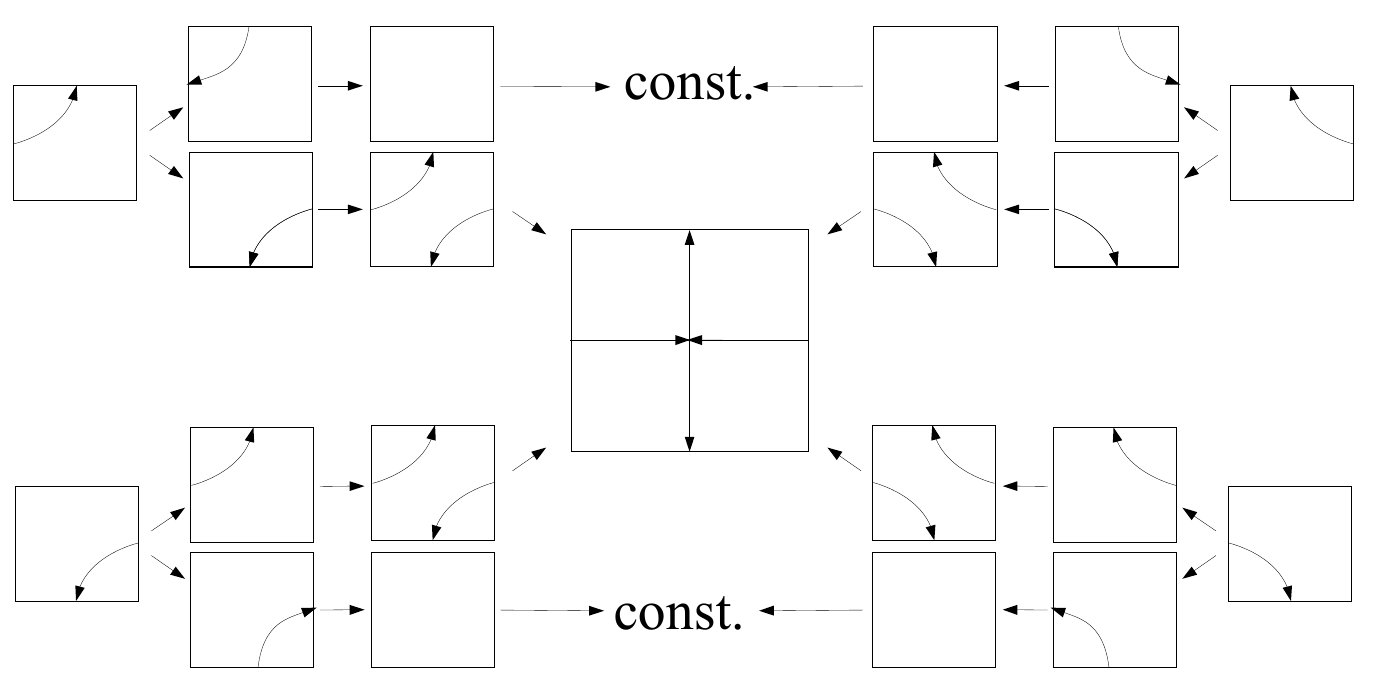}\caption{Generation of plaquette interactions as effective second-order terms,
as in \cite{Zohar2011}. First, an atom hops from a link
to a neighboring one (diagonally). Then the Gauss constraint (\ref{eq:HG})
is violated, and a second process, restoring the constraint, must
take place. This may either be another hopping process, with other
two links, closing a plaquette, or the inverse hopping which results
in {}``canceling'' the first one, resulting in a constant term in
the approach of \cite{Zohar2011} (and in other gauge-invariant terms
in the spin-gauge approach \cite{Zohar2012}, where the link operators
are not unitary). Each plaquette is constructed out of four different
second order processes. Note that the arrows correspond to the {}``modified
Gauss's law'' (\ref{eq:modgauss}), with opposite signs on half of
the links - i.e. to the language of the simulating systems (atoms)
and not to the {}``right signs'' of $cQED$, obtained after the
canonical transformation (\ref{eq:cantrans}).\label{fig:Generation-of-plaquette}}

\par\end{centering}

\end{figure}

Neighboring links are populated by condensates of different atomic
species, and thus direct, gauge-variant tunneling is eliminated. This
is possible by the use of holographic masks, for example, for the creation
of the optical lattice. Local scattering processes are responsible to
the electric Hamiltonian, as well as, along with neighboring links
scattering, Gauss's law constraint. Using external lasers, desired
hopping processes between neighboring links (diagonally) are introduced (see figure \ref{BECfig}).
These terms are gauge-variant as well, but they induce effective gauge
invariance: these hopping processes correspond to violation of the
constraint, but second-order perturbation theory ties pairs of them
together, forming the plaquette interactions (see figure \ref{fig:Generation-of-plaquette}).

Experimental considerations (homogeneity of the scattering lengths)
result in a sum-Gauss's law, rather than with a divergence, i.e.,
\begin{equation}
G_{m,n}=\delta{}_{m,n}^{1}+\delta{}_{m,n}^{2}+\delta{}_{m-1,n}^{1}+\delta{}_{m,n-1}^{2}-\Delta_{m,n}\label{eq:modgauss}
\end{equation}
and thus a (theoretical) canonical transformation, involving inversion
of signs, must be carried out in order to obtain the {}``right''
$cQED$ degrees of freedom,
\begin{equation}
E{}_{m,n}^{k}=\left(-1\right)^{m+n}\delta{}_{m,n}^{k};\theta{}_{m,n}^{k}\longrightarrow\left(-1\right)^{m+n}\theta{}_{m,n}^{k};q{}_{m,n}=\left(-1\right)^{m+n}\Delta_{m,n}\label{eq:cantrans}
\end{equation}
Being theoretical, this transformation should be carried out on the
results of the measurements: a strong coupling electric flux-tube
will thus be manifest by a line of \emph{alternating} values of $\delta{}_{m,n}^{k}$
between the charges.

This proposal also allows leaving the perturbative (in terms of the
simulated theory) strong-coupling regime, but to a limited extent, since
the atomic excess must be small comparing to the mean population,
which cannot be too large due to experimental consequences. However,
in 2+1 dimensions, where confinement holds also in the weak limit,
the simulation of the
weak limit should work - at least qualitatively, even if not describing
the exact weak limit of the full Kogut-Susskind Hamiltonian. This
can be understood from the weak limit analysis of \cite{Zohar2012}.

\subsection{Truncated cQED model - link interactions}

The second proposal \cite{Zohar2012} was a first step in the direction of a simpler
implementation. Instead of Bose-Einstein condensates, each link is
occupied by a single atom, belonging to each of $2\ell+1$ possible
atomic species, representing $2\ell+1$ possible values of the electric
field on a link: Here we have introduced the \emph{spin-gauge} model,
in which the electric field is truncated (taking values between $-\ell$
and $\ell$), but the gauge symmetry is unaffected, which should lead,
at least qualitatively, to similar effects those of as the Kogut-Susskind Hamiltonian.
The $U\left(1\right)$ angular momenta are replaced by $SU\left(2\right)$
angular momenta in an integer representation $\ell$ with $2\ell+1$
levels, with the $L$ operators of the electric field replaced by
the truncated $L_{z}$, and the unitary raising/lowering phase operators
replaced by the non-unitary $SU\left(2\right)$ ladder operators $L_{\pm}$:
\begin{equation}
U\longrightarrow\frac{1}{\sqrt{\ell\left(\ell+1\right)}}L_{+}
\end{equation}
in a confining phase, the electric field is supposed to take small
values $m$, and thus with a large enough $\ell$, as well as the
$m\ll\ell$ of confinement, the unitarity of the ladder operators
is approximately restored, at least in the sense that when one acts with them
 on the $\left|\ell m\right\rangle$ states, no prefactor is obtained:
 \begin{equation}
\frac{1}{\sqrt{\ell\left(\ell+1\right)}}L_{\pm}\left|\ell m\right\rangle =\sqrt{1-\frac{m\left(m\pm1\right)}{\ell\left(\ell+1\right)}}\left|\ell,m\pm1\right\rangle \approx\left|\ell,m\pm1\right\rangle
\end{equation}

In this representation, one obtains the \emph{Spin-Gauge Hamiltonian} \cite{Zohar2012},
\begin{equation}
H_{SG}=\frac{g^{2}}{2}\underset{\mathbf{n},k}{\sum}L_{z,\mathbf{\mathbf{n},}k}^{2}-\frac{1}{2g^{2}\ell^{2}\left(\ell+1\right)^{2}}\underset{plaquettes}{\sum}\left(L_{+,\mathbf{\mathbf{n},}1}L_{+,\mathbf{\mathbf{n+\hat{1}},}2}L_{-,\mathbf{\mathbf{n+\hat{2}},}1}L_{-,\mathbf{\mathbf{n},}2}+H.c.\right)\label{eq:HSG}
\end{equation}

We have compared this model to the Kogut-Susskind Hamiltonian (\ref{eq:abKS}).
First, we have shown the more obvious result, that in the strong limit,
as the states are eigenstates of electric flux (see subsection \ref{sub:Strong-Coupling-Lattice}),
one may use a truncated theory which suffices to describe the state
of the system in the presence of static charges. Conditions on $\ell$,
depending on the magnitude of the static charges, have been discussed
using perturbation theory considerations.
It was shown what is the highest order in strong-limit perturbation
theory, in which the spin-gauge states will correspond to the Kogut-Susskind
ones, for a given charge configuration.

Furthermore, we have addressed the nonperturbative, nontrivial weak
limit as well, and compared numerically the single-plaquette results
for a truncated Kogut-Susskind theory as well as the spin gauge theory
to the analytical results of \cite{DrellQuinnSvetitskyWeinstein}.
The results converge quantitatively to the analytical ones very quickly
- for small values of $\ell$ (already for $\ell\sim2,3$).

An implementation for $\ell=1$ - three bosonic internal levels -
has been introduced. It involves a three-scale hierarchy - the first, strongest
constraint is responsible to the effective "angular-momentum"
interactions \cite{Mazza2010} and the second one to gauge invariance,
in a similar manner to \cite{Zohar2011}.
Working with $\ell=1$ limits the possibility of going into the weak
limit. Nevertheless, the rapid convergence of the results in the weak
limit to the ones of the Kogut-Susskind model, do show that implementations
of the spin-gauge model for higher values of $\ell$ shall serve as
an accurate weak-limit simulation. Confinement and flux tubes take
place here in the form of alternating atomic species along the line
connecting the static charges - in the strong limit.

Due to the non-unitarity of the $L_{\pm}$ operators, the effective
constructions involves some undesired terms. However, they are gauge-invariant
as well (as they satisfy Gauss's law constraint) and their effect
is discussed in \cite{Zohar2012}.

Finally, in \cite{Zohar2013}, we have introduced dynamical fermions as well. Still considering $cQED$, in the spin-gauge model presented
in paper \cite{Zohar2012}, but regardless of the implementation the
gauge degrees of freedom, we have introduced naive fermions into the
system, in 2+1 dimensions.

Working with naive fermions in two spatial dimensions, each vertex
contains two fermionic species, simulating a 2-component Dirac spinor,
\begin{equation}
\psi_{\mathbf{n}}=\left(\begin{array}{c}
c_{\mathbf{n}}\\
d_{\mathbf{n}}
\end{array}\right)
\end{equation}
with the local charge
\begin{equation}
Q_{\mathbf{n}}=\psi_{\mathbf{n}}^{\dagger}\psi_{\mathbf{n}}-1=N_{\mathbf{n}}^{C}+N_{\mathbf{n}}^{D}-1
\end{equation}
(similarly to the charge in the second quantized Dirac field. There,
for the lower entry of the spinor, we swap the fermionic creation
and annihilation operators, and thus obtain that this charge transforms
to the well known $N_{\mathbf{n}}^{C}-N_{\mathbf{n}}^{D}$ form).

The simulated Hamiltonian (besides the pure-gauge part, (\ref{eq:HSG}))
in 2+1 dimensions, is
\begin{equation}
H_{naive}=\frac{i\epsilon}{\sqrt{\ell\left(\ell+1\right)}}\underset{\mathbf{n},k}{\sum}\left(\psi_{\mathbf{n}}^{\dagger}\sigma_{k}\psi_{\mathbf{n+\hat{k}}}L_{+,\mathbf{\mathbf{n},}k}-H.c.\right)+M\underset{\mathbf{n},k}{\sum}\psi_{\mathbf{n}}^{\dagger}\sigma_{z}\psi_{\mathbf{n}}
\end{equation}
with the Dirac matrices being Pauli matrices, $\alpha_{1,2}=\sigma_{x,y}$
and $\beta=\sigma_{z}$. The fermionic mass terms manifest explicitly
that the particles and antiparticles are not divided among vertices
as in the staggered formulation.

Gauge invariance is still effective here. Inclusion of dynamical charges
into the Gauss's law constraints made it harder to implement, requiring
also alternating signs to the dynamical charges. A detailed prescription
for the implementation, based on fermion-boson scattering, was given,
enabling the generation of this required constraint - the interaction
between the gauge field and the matter arises as an effective second
order term as well: in order to satisfy the constraint, fermions can only hop
along with a change of the electric field on the link.

Generalization of this method to other types of lattice fermions is straightforward. One simply has to change
the fermionic ingredients and tailor the desired hopping. The effective construction will "make sure" that only
the gauge-invariant interactions are obtained.

\subsection{Effective Gauge Invariance - some general remarks}

The three proposals \cite{Zohar2011,Zohar2012,Zohar2013} involved
gauge invariance as an effective symmetry. This allows introducing
gauge invariance to the simulating systems, but it is not the only possible
approach, as shown in the subsequent papers \cite{AngMom,NA}.

The Gauss's law constraint involves scattering of atoms,
representing the gauge field degrees of freedom, on several neighboring
links. All the interactions must be with the same amplitude, which
is hard to implement. This may lead to practical impossibilities to
satisfy the Gauss's law, for example, or to the unintended inclusion
of fractional charges.

The robustness of this method was considered by Kasamatsu et. al.
\cite{Kasamatsu2013,Kuno2014}, who showed, mainly based on \cite{Zohar2011}
but also on other works of us and the other groups, that one may relax
the fine-tuning demands of the simulating Hamiltonian parameters,
and still get a gauge-invariant theory, involving Higgs fields.

Another disadvantage of this approach applies to simulation of non-Abelian
gauge theories. There, the generators of gauge transformations, required
for the Gauss's law constraint, are much more complicated (several
non-commuting generators (\ref{eq:Gaugegen})), which makes the constraint
Hamiltonian (\ref{eq:HG}) very difficult to realize.

\section{Local gauge invariance arising from many body interaction symmetries}

The basic idea of the second simulation approach we describe hereby, is that one could use local
interactions of the many body system - atomic collisions - to achieve local gauge invariance.
In particular, we have utilized the
conservation of the hyperfine angular momentum in atomic collisions.
This enables to obtain, as fundamental ingredients of the
Hamiltonian, the on-link gauge-matter (boson-fermion) interactions of (\ref{eq:HGM}), without any constraints or perturbation theory \cite{AngMom,NA}.
As a second step, in $2+1$ (and more) dimensional setups, one may introduce the \emph{loop method} \cite{AngMom} presented below,
which allows the construction of the plaquette interactions effectively, but out of the \emph{already gauge invariant} on-link interactions.

The basic, on-link interactions (\ref{eq:HGM}) are obtained from
boson-fermion collisions, utilizing the overlap of four Wannier functions
on a link (two bosonic on the links and two fermionic on its edges).
The only allowed scattering channels are those which conserve the
hyperfine angular momentum. Thus, one can specifically choose the
hyperfine levels playing the roles of the different \emph{simulated}
bosons and fermions, such that the $F$ conserving processes will
be mapped to the gauge invariant ones. This gives us the desired interactions,
along with other ones, gauge invariant as well, which may be tolerated
or tamed.

\subsection{Compact QED - on-link interactions}

We shall show this procedure in detail for compact $QED$ in 1+1 dimensions -
the lattice Schwinger model.
Besides its advantage as a "toy model" which allows a clear and simple presentation of the method,
the quantum simulation of 1+1 dimensional $QED$ with matter, in the manner presented below, has several advantages:
\begin{enumerate}
\item As gauge invariance is fundamental here, no use of effective interactions
is required - i.e., the simulating Hamiltonian produces the simulated
theory without any use of perturbation theory. This implies that the energy scales are larger,
and thus the simulation should be faster and more robust to noise and decoherence.
\item It is a possibly simpler implementation (compared with \cite{Zohar2011}
and \cite{Zohar2012}).
\item The 1+1 dimensional Schwinger model is exactly solvable \cite{Schwinger1962I,Schwinger1962}.
Lattice results are available as well (\cite{Banks1976} and much
more, including new calculations involving matrix product states \cite{Banuls2013}).
Along with 3, this make this model a possibly preferred model for
a first physical realization, as one has analytical and numerical
results to compare with.
\end{enumerate}

\begin{figure}[t]
\begin{centering}
\includegraphics[scale=0.6]{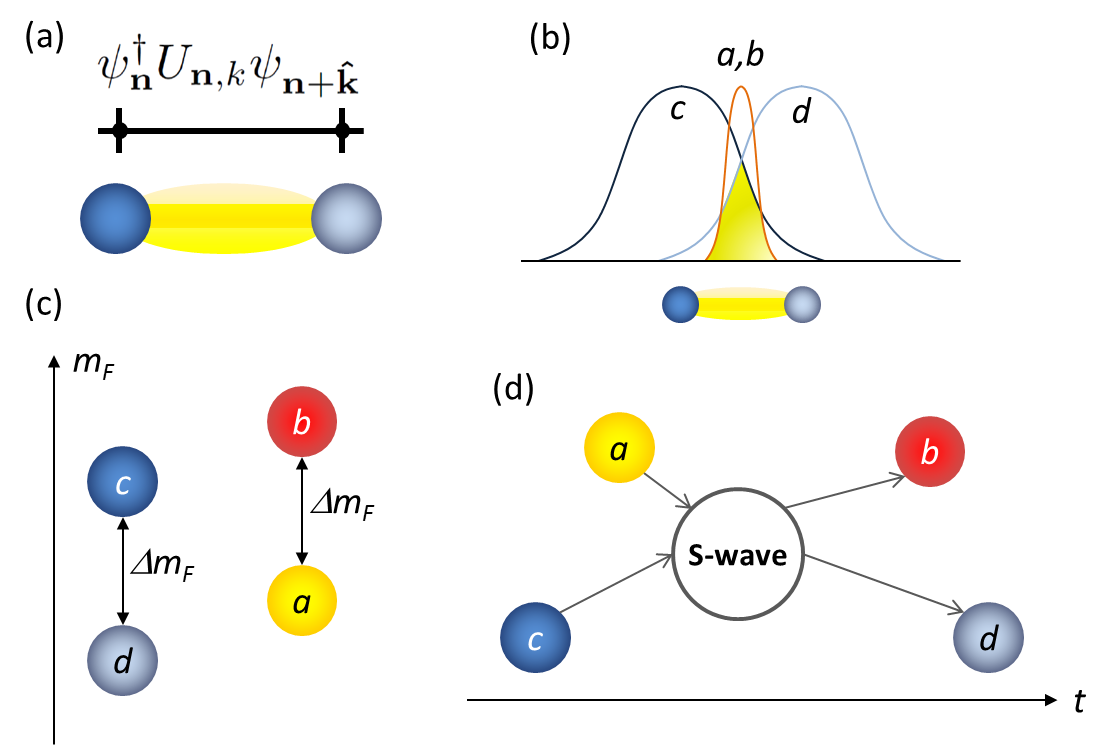}
\par\end{centering}

\caption{"Exact gauge invariance" - the Schwinger model example. (a) The desired interactions (\ref{eq:HGM})
are obtained as the $F$ preserving boson-fermion scattering processes,
utilizing the overlap of bosonic and fermionic Wannier functions on
the link (b). By appropriately choosing the hyperfine levels representing
the bosons and the fermions (c), only the gauge invariant interactions
(\ref{eq:gip}) of (\ref{eq:Hsc}) are possible (d). }

\label{fig:Exact-gauge-invariance.}
\end{figure}

While the fermions are simply fermionic atoms in this simulation (staggered
ones, to be specific, but they may be generalized to other lattice
fermion representations), the bosonic fields are constructed out of
the so-called \emph{Schwinger bosons}, utilizing Schwinger's representation
of $SU(2)$ \cite{Jordan,Schwinger}:
In this $cQED$ simulation, each link may be populated by two different
bosonic species, $a$ and $b$, from which the algebra is constructed
in the following way:
\begin{equation}
L_{z}=\frac{1}{2}\left(a^{\dagger}a-b^{\dagger}b\right)\quad;\quad\ell=\frac{1}{2}\left(a^{\dagger}a+b^{\dagger}b\right)\label{eq:schinw}
\end{equation}
\begin{equation}
L_{+}=a^{\dagger}b\quad;\quad L_{-}=b^{\dagger}a
\end{equation}
using this representation, along with the conservation of hyperfine
angular momentum, simulations of the spin-gauge Hamiltonian may be
obtained, but much simply than in the effective approach. Assuming
that the bosons do not interact among links, $\ell$ is fixed and
governed by the number of on-link bosons. If we consider a single
link, whose left edge may be occupied only by the fermionic species
$c$ , and its right edge by $d$, we wish to obtain the interactions
\begin{equation}
c^{\dagger}L_{+}d+d^{\dagger}L_{-}c=c^{\dagger}a^{\dagger}bd+d^{\dagger}b^{\dagger}ac\label{eq:gip}
\end{equation}
and this is possible by selecting
\begin{equation}
m_{F}\left(a\right)+m_{F}\left(c\right)=m_{F}\left(b\right)+m_{F}\left(d\right)
\end{equation}
(see figure \ref{fig:Exact-gauge-invariance.}). Even two atoms per
link, equivalent to $\ell=1$, give rise to the required gauge symmetry.
But enlarging $\ell$ here involves many atoms of the same two species
- condensates - rather than many degrees of freedom, as in \cite{Zohar2012}.
The electric energy term of $L_{z}^{2}$ is obtained by other {}``gauge
invariant'' scattering processes, between the bosons of a single
link.

This simulation proposal has been treated numerically, using matrix product states \cite{Kuhn2014}.

\subsection{$SU(2)$ - Yang Mills theory - on-link interactions}

\subsubsection{Left and right Schwinger representations.}
One can utilize this type of interactions for other gauge theories as well,
such as the $SU(2)$ lattice gauge theory. In \cite{NA}, we have
suggested a way for realizing an $SU(2)$ lattice gauge
theory, in $1+1$ dimensions, utilizing the prepotential method \cite{MathurSU2,Mathur2006,Mathur2007,Anishetty}
for representing the gauge degrees of freedom (which generalizes the
Schwinger representation). In this method, every link is decomposed
into two parts - left and right, corresponding the left and right
degrees of freedom. The Hilbert space
of each edge of the link consists of two harmonic oscillators, generating
a separate Schwinger algebra: $a_{1,2}$ on the left side,  with the left $SU(2)$ algebra (\ref{eq:alg3}) generated by
\begin{equation}
L_{\alpha} = \frac{1}{2}\underset{ij}{\sum} a^{\dagger}_{i}\left(\sigma_{\alpha}\right)_{ji}a_{j}
\end{equation}
and $b_{1,2}$ on the right side,  with the right $SU(2)$ (\ref{eq:alg4}) algebra generated by
\begin{equation}
R_{\alpha} = \frac{1}{2}\underset{ij}{\sum} b^{\dagger}_{i}\left(\sigma_{\alpha}\right)_{ij}b_{j}
\end{equation}
The left and right couples of oscillators have the same total number of excitations -
\begin{equation}
N_L \equiv a^{\dagger}_{1}a_{1} + a^{\dagger}_{2}a_{2} = b^{\dagger}_{1}b_{1} + b^{\dagger}_{2}b_{2} \equiv N_R
\end{equation}
(which is proportional to the total angular momentum in the Schwinger
representation (\ref{eq:schinw})), and thus, relation (\ref{eq:SU2casimir})
is satisfied and the correct Hilbert space, with three degrees of
freedom and left and right $SU(2)$ algebras, is achieved. The rotation
matrices' elements are constructed out of creation and annihilation
operators of these oscillators. One can decompose them to $U = U_L U_R$, with the matrices
\begin{equation}
U_L = \frac{1}{\sqrt{N_L + 1}}\left(
        \begin{array}{cc}
          a_1^{\dagger} & -a_2 \\
          a_2^{\dagger} & a_1 \\
        \end{array}
      \right)
 \; ; \;
U_R = \left(
        \begin{array}{cc}
          b_1^{\dagger} & b_2^{\dagger} \\
          -b_2 & b_1 \\
        \end{array}
      \right)
      \frac{1}{\sqrt{N_R + 1}}
\end{equation}
defined on both sides of the link.

\begin{figure}[t]
\begin{centering}
\includegraphics{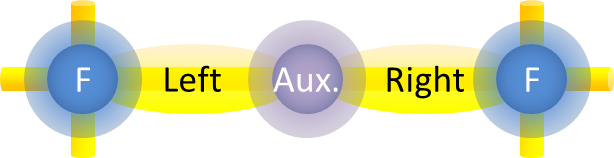}
\par\end{centering}
\caption{Structure of the non-Abelian lattice atomic simulator in the prepotential method: the simulated link is decomposed into two pieces, tied together by an auxiliary fermion. }
\label{nonab}
\end{figure}

The implementation presented in \cite{NA} involves bosons on the links and fermions on the
vertices, as before, but now, due to the new, exact gauge invariance
method, the hyperfine levels of the atoms must be chosen carefully,
such that only the desired scattering processes, which correspond
to gauge invariant interaction, will be allowed by hyperfine angular
momentum conservation. The links were not established as "first
principle interactions", but rather "glued" out of two building blocks,
using auxiliary, ancillary fermions, in second order perturbation
theory, forming effective link interactions: each of the building blocks is obtained using the above prescription,
involving transition to the auxiliary fermionic site in the middle, whose energy is too high, and in order to satisfy this
constraint, one obtains, in second order, the full $SU(2)$ link (see figure \ref{nonab}).
Nevertheless, gauge invariance is not effective in this proposal in
the sense there is no Gauss's law constraint, and this is the achievement
of this work. The link bosons involved contain four species playing
the role of the harmonic oscillators of the prepotential method, from
which the algebras are constructed, and other four "auxiliary"
species, serving as "reference baths" required for the number-conserving
scattering processes.

However, this is not a simulation of the "full"
$SU(2)$ Kogut-Susskind theory using the exact prepotential Hamiltonian;
Much like some of the $cQED$ simulations (in the spin gauge method),
in which the electric field was truncated, here it was restricted
to the representations $j=0,\frac{1}{2}$. This, along with the square
roots in the denominators of prepotential representation of the group elements (rotation matrices)
- that were not achieved in the simulation - have generated some inaccuracies
compared to the Kogut-Susskind theory, which result in the dynamics
being accurate only in a regime where one may consider the gauge-matter
interactions as a small perturbation (to sixth order in it).

\subsubsection{Truncation scheme for the Hilbert space of non-Abelian groups.}
One can overcome the problem of square roots of operators in the denominator,
as well as the need for auxiliary bosonic species and decomposition of the links,
if the truncation scheme of \cite{PRD} is utilized. In this method,
we still work with a truncated theory, but the truncation
is done in a gauge invariant way, which does not involve decomposition of the
link into two (left and right) pieces.

The bosonic degrees of freedom are represented by single bosonic atoms residing on each link
of the lattice. The state of each such atom may be one of $\mathcal{N}=\underset{j=0}{\overset{J_{max}}{\sum}}\left(2j+1\right)^2$ internal levels,
labeled by $\left|j m m'\right\rangle$.
These are generated from the local atomic vacuum state $\left|0\right\rangle$ by the creation operators
$a^{\dagger j}_{mm'}$, with $j \in \left\{q/2\right\}_{q=0}^{2J_{max}}$, $-j \leq m,m' \leq j$:
\begin{equation}
\left|j m m'\right\rangle = a^{\dagger j}_{mm'} \left|0\right\rangle
\end{equation}

Out of these creation operators, and their conjugate annihilation operators, one can construct the $SU(2)$ algebras
\begin{equation}
\begin{aligned}
L_{\alpha}  = \underset{j}{\sum} a^{\dagger j}_{mm'} \left(T_{\alpha}^{j}\right)_{nm} a^{j}_{nm'} &&
R_{\alpha}  = \underset{j}{\sum} a^{\dagger j}_{mm'} \left(T_{\alpha}^{j}\right)_{m'n'} a^{j}_{mn'}
\end{aligned}
\end{equation}
where a summation is assumed on double indices. $T_{\alpha}^{j}$ are the $j$th representation matrices of $SU(2)$ (for example, $T_{\alpha}^{1/2} = \sigma_{\alpha}/2$).
These operators satisfy the left and right $SU(2)$ algebras presented above.
Ladder operators may be defined as $L_{\pm} = L_{x} \mp  i L_{y} = \underset{j}{\sum} a^{\dagger j}_{mm'} \left(T_{\pm}^{j}\right)_{mn} a^{j}_{nm'}$  and $R_{\pm} = R_{x} \pm i R_{y} = \underset{j}{\sum} a^{\dagger j}_{mm'} \left(T_{\pm}^{j}\right)_{m'n'} a^{j}_{mn'}$.
All these operators satisfy the algebras

The commutation relations with the creation operators are
\begin{equation}
\begin{aligned}
\left[L_{\alpha},a^{\dagger j}_{mm'}\right]  & = \left(T_{\alpha}^{j}\right)_{mn}a^{\dagger j}_{nm'} \\
\left[R_{\alpha},a^{\dagger j}_{mm'}\right]  & = a^{\dagger j}_{mn'} \left(T_{\alpha}^{j}\right)_{n'm'}
\end{aligned}
\end{equation}
implying that they undergo left and right rotations within the $j$ representation. Thus, if we perform a rotation $V \in SU(2)$ on the left and $W^{\dagger}\in SU(2)$ on the right, these operators will transform according to
\begin{equation}
a^{\dagger j}_{mm'} \rightarrow V^{j}_{mn} a^{\dagger j}_{nn'} W^{\dagger j}_{n'm'}
\label{opgauge}
\end{equation}
with $V^{j},W^{\dagger j}$ the $j$th matrix  representation of the group elements $V,W^{\dagger}$.

Next, one would like to exploit these transformation properties for the construction of rotation matrices - i.e., $SU(2)$ group elements.
This is done, as explained for general groups in \cite{PRD}, using the Clebsch-Gordan series.
We denote the Clebsch-Gordan coefficients by $\alpha^{jMN}_m \left(J,K\right)= \left\langle J j M m | K N \right\rangle$ , and define
\begin{equation}
u^{j}_{mm'}\left(J,K\right) = \alpha^{jMN}_m \left(J,K\right)  \alpha^{jM'N'}_{m'} \left(J,K\right) a^{\dagger K}_{NN'} a^{J}_{MM'}
\end{equation}
The $u^{j}_{mm'}\left(J,K\right)$ operators undergo rotations according to the $j$th representation, and
thus the corresponding rotation matrix elements \cite{Rose1995} may be constructed out of them:
\begin{equation}
U^{j}_{mm'} = \underset{J}{\sum}\underset{K=|J-j|}{\overset{J+j}{\sum}}\sqrt{\frac{2J+1}{2K+1}} u^{j}_{mm'}\left(J,K\right)
\label{USum}
\end{equation}
 However, as the $u^{j}_{mm'}\left(J,K\right)$ operators undergo rotations independently of each other, we can also construct objects with the desired transformation properties without including an infinite number of levels, i.e. with $J_{max}<\infty$. This is the result of truncating the sums in the generators and the rotation matrices, limiting them to $J \leq J_{max}$.
Thus, also for a finite $J_{max}$ one can obtain rotation matrices operating on the $j<J_{max}$ sectors, and this allows the construction of a \emph{truncated} but yet \emph{gauge invariant} lattice gauge theory, as we shall describe next. The unitarity is lost, and now
\begin{equation}
\text{tr}\left(U^{j\dagger}U^j\right) = \text{tr}\left(U^jU^{j\dagger}\right) = 2j+1 - f_j\left(J_{max}\right)P_{J_{max}}
\label{trace}
\end{equation}
where $P_{j} = \underset{m,m'}{\sum}a^{\dagger j}_{mm'}a^{j}_{mm'}$ is a projection operator to the $j$ sector (recall that we are dealing with the Hilbert subspace of a single atom).

The electric Hamiltonian will simply consist of $P_j$ projectors with the right prefactors, corresponding to linear terms in the bosonic number operators. The gauge-matter interactions are, however, more complicated.

\begin{figure}[t]
\begin{centering}
\includegraphics{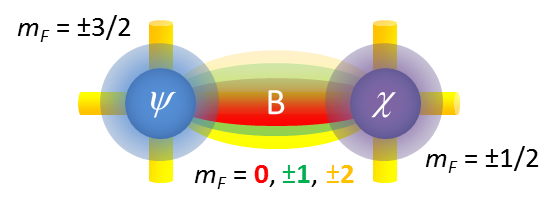}
\caption{The simulating lattice structure. As usual, the bosonic potential's minima are on the links, and the fermionic - on the vertices.
The different shapes of the bosonic minima are schematically presented.}
\label{fig9}
\end{centering}
\end{figure}

\subsubsection{Realization of the truncated on-link interactions.}
Next we shall give an example for the realization of the above truncation, with the $j=0,1/2$ representations.

The scheme we were studying contains bosons belonging to a $F=2$ hyperfine manifold, with five different hyperfine levels (as required for $J_{max}=\frac{1}{2}$).
They are prepared such that each link is populated, as required for our representation, by a single atom, and the bosonic optical
 wells are deep enough, such that no hopping of bosons between different bosonic sites are possible. Thus the system remains
 in the local single-atom Hilbert subspace. The bosonic creation operators are $\left\{b^{\dagger}_m\right\}$,
 for $-2 \leq m \equiv m_F \leq 2$. The trapping potential's minima of the different bosonic species have different shapes (generated by different internal potentials) -
 there are at least three different such shapes, for $|m|=0,1,2$ (see figure \ref{fig9}).
  The fermions belong to an $F=3/2$ manifold, with four different hyperfine levels. These are required for the mapping between $F$ conservation in atomic collisions and gauge invariance.
      Each vertex (fermionic site) may be occupied by two fermions at most, and in an alternating manner: the even sites may be occupied by $m_F = \pm 3/2$ fermions, while the odd ones by $m_F = \pm 1/2$ only. We label even vertices' fermions by $\psi$ and the odd by $\chi$, where $m_F \left( \psi_1 \right) = -3/2$, $m_F \left( \psi_2 \right) = 3/2$,
$m_F \left( \chi_1 \right) = 1/2$, and $m_F \left( \chi_2 \right) = -1/2$.
     The fermionic sites are designed such that the fermionic wavefunctions of the two edges of a link will
      overlap on the middle on the link, also with the bosonic wave functions (see figure \ref{fig9}).

\begin{figure}[t]
\begin{centering}
\includegraphics[scale=0.6]{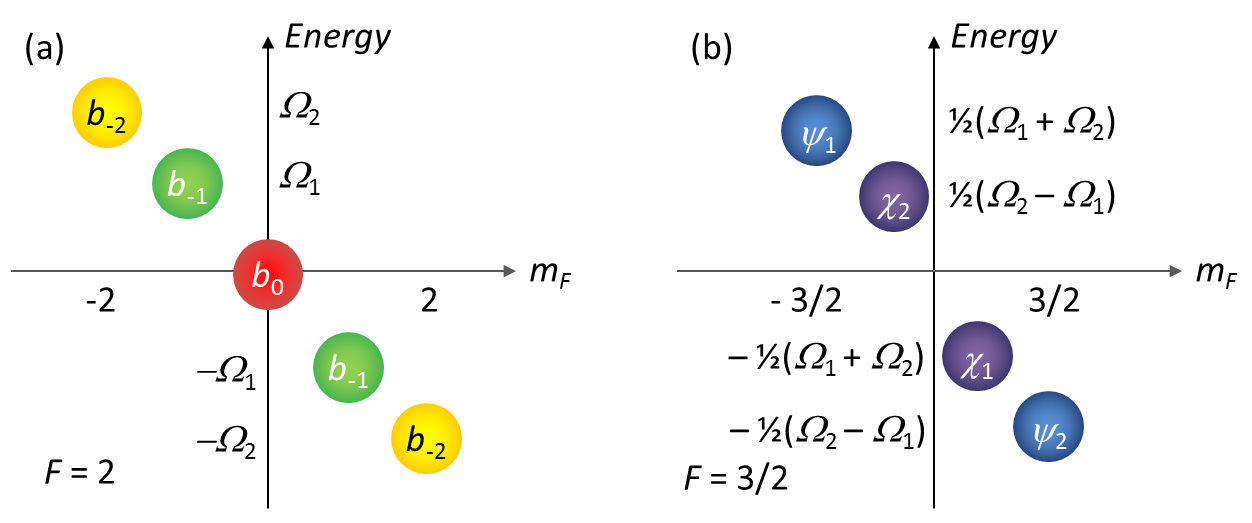}
\caption{The energy levels of the (a) $F=2$ bosons  and the (b) $F=3/2$ fermions , as described by $H_{\Omega}$.}
\label{fig2}
\end{centering}
\end{figure}

Using lasers or magnetic fields, depending on the atomic realization, the bosons and fermions are subject to the Hamiltonian
\begin{equation}
\begin{aligned}
H_{\Omega} & = -\underset{L}{\sum} \underset{m \neq 0}{\sum} \Omega_{\left|m\right|} b^{\dagger}_m b_m + \frac{1}{2} \underset{V_e}{\sum}\left(\Omega_1 + \Omega_2\right)\left( \psi^{\dagger}_1 \psi_1 - \psi^{\dagger}_2 \psi_2\right) \\
& + \frac{1}{2} \underset{V_o}{\sum}\left(\Omega_1 - \Omega_2\right)\left( \chi^{\dagger}_1 \chi_1 - \chi^{\dagger}_2 \chi_2\right)
\end{aligned}
\end{equation}
where $L$ is the set of links, $V_e$ of even vertices and $V_o$ of odd ones, and $\Omega_1 \neq \Omega_2$, $2\Omega_1 \neq \Omega_2$
(see figure \ref{fig2}).

Gauge invariant interactions, according to (\ref{eq:HGM}), are of the form $\psi^{\dagger} U \chi$, where $U$ is the group element, which is a matrix of bosonic operators, as given by (\ref{USum}) with $j=1/2$ and $J_{max}=1/2$.  These interactions are constructed by the on-link fermion-boson scattering. The chosen values of $m_F$ allow only specific scattering channels, thanks to the conservation of hyperfine angular momentum. By the introduction of $H_{\Omega}$ and energy conservation considerations, we further reduce the number of possible scattering processes (assume $\Omega_{1,2}$ are large enough compared to the other energy scales, and perform a rotating wave approximation with respect to $H_{\Omega}$). In first quantization terms, the S-wave scattering is governed by a pseudo-potential $V_S$ whose matrix elements are
\begin{equation}
\begin{aligned}
\left\langle 2,m'_{b},\frac{3}{2},m'_{f}\right|V_{S}\left|2,m_{b},\frac{3}{2},m_{f}\right\rangle & = \\
\delta_{m'_{b}+m'_{f},m{}_{b}+m{}_{f}}\overset{\frac{7}{2}}{\underset{F=\frac{1}{2}}{\sum}}C_{F} \times \left\langle 2,m'_{b},\frac{3}{2},m'_{f}|F,m'_{b}+m'_{f}\right\rangle & \times \\ \times\left\langle F,m_{b}+m_{f}|2,m_{b},\frac{3}{2},m_{f}\right\rangle
\end{aligned}
\label{VS}
\end{equation}
(The Kronecker delta is redundant due to the Clebsch-Gordan coefficients, but is explicitly written to manifest angular momentum conservation.)

By properly choosing the $C_F$ coefficients and the overlap integrals (by manipulating the shape of the optical potential and hence the Wannier functions) one may obtain the required link interactions as in (\ref{USum}) and the electric energy as in equation (\ref{HE}) (which is simply a sum of bosonic number operators, using the projectors $P_j$ defined above), as well as avoid extra interaction terms which conserve hyperfine angular momentum, but yet are not part of the desired Hamiltonian.
Unfortunately, however, we have not been able, so far, either to satisfy all the required symmetries by the simulating system, or to prove it is impossible.
An experimental realization might be possible for other atomic configurations, with less stringent symmetries (i.e., a different choice of atomic levels).

For the simplest configuration described above (which could be generalized to more complicated atomic systems if required),
with a theoretical simulating system with the right Hamiltonian parameters, one could proceed as follows:
For even links we obtain the interaction Hamiltonian
\begin{equation}
\epsilon\underset{ij}{\sum}\psi^{\dagger}_i M_{ij} \chi_j+ H.c.
\end{equation}
and for odd ones -
\begin{equation}
\epsilon\underset{ij}{\sum}\chi^{\dagger}_i M^{\dagger}_{ij} \psi_j+ H.c.
\end{equation}
with the bosonic matrix
\begin{equation}
M=\frac{1}{\sqrt{2}}\left(\begin{array}{cc}
b_{2}^{\dagger}b_{0}+b_{0}^{\dagger}b_{-2} & -b_{1}^{\dagger}b_{0}+b_{0}^{\dagger}b_{-1}\\
-b_{-1}^{\dagger}b_{0}+b_{0}^{\dagger}b_{1} & b_{0}^{\dagger}b_{2}+b_{-2}^{\dagger}b_{0}
\end{array}\right)
\end{equation}

The next step is to map between the atomic degrees of freedom $b_m$ to the ones of the simulated theory - $a^{j}_{mm'}$. The mapping is as follows:
\begin{enumerate}
  \item For even links,
  \begin{equation}
  \begin{aligned}
  b^{\dagger}_{\pm 2} &= a^{\dagger 1/2}_{\pm 1/2,\pm 1/2} \\ b^{\dagger}_{\pm 1} &= -a^{\dagger 1/2}_{\pm 1/2,\mp 1/2} \\ b^{\dagger}_{0} &= a^{\dagger 0}_{00}
  \end{aligned}
  \label{M1}
  \end{equation}
  \item For odd links,
  \begin{equation}
  \begin{aligned}
  b^{\dagger}_{\pm 2} &= a^{\dagger 1/2}_{\mp 1/2,\mp 1/2} \\ b^{\dagger}_{\pm 1} &= a^{\dagger 1/2}_{\pm 1/2,\mp 1/2} \\ b^{\dagger}_{0} &= a^{\dagger 0}_{00}
  \end{aligned}
  \label{M2}
  \end{equation}
\end{enumerate}
Using this mapping, one obtains that $M=U$ for even links, and $M^{\dagger}=U$ for the odd ones (with $U$ being the truncated rotation matrix, acting only within the 5-dimensional Hilbert
space of $j=0,\frac{1}{2}$.

Thus, if we relabel all the fermions as $\Psi$, we obtain the desired interaction Hamiltonian,
\begin{equation}
H_{int}=\frac{\epsilon}{\sqrt{2}}\overset{}{\underset{n}{\sum}}\left(\Psi_{n}^{\dagger}U_{n,n+1}\Psi_{n+1}+H.C.\right)
\end{equation}

\subsection{Plaquette interactions: $d+1$ dimensional Abelian and non-Abelian theories }
The non-effective simulations may be extended to further dimensions. The nontrivial
generalization is from 1+1 to 2+1 dimensions, as it involves the inclusion
of the plaquette interactions. In the previous methods, which involved
effective gauge invariance, these were effectively obtained out of
the \emph{gauge-variant} interactions of the original Hamiltonian,
using the gauge invariance constraint. Here they are obtained effectively
as well, but out of \emph{already gauge invariant} building blocks
- the on-link gauge-matter interactions, which are not effective here,
which is a great advantage.

\begin{figure}[t]
\begin{centering}
\includegraphics[scale=0.5]{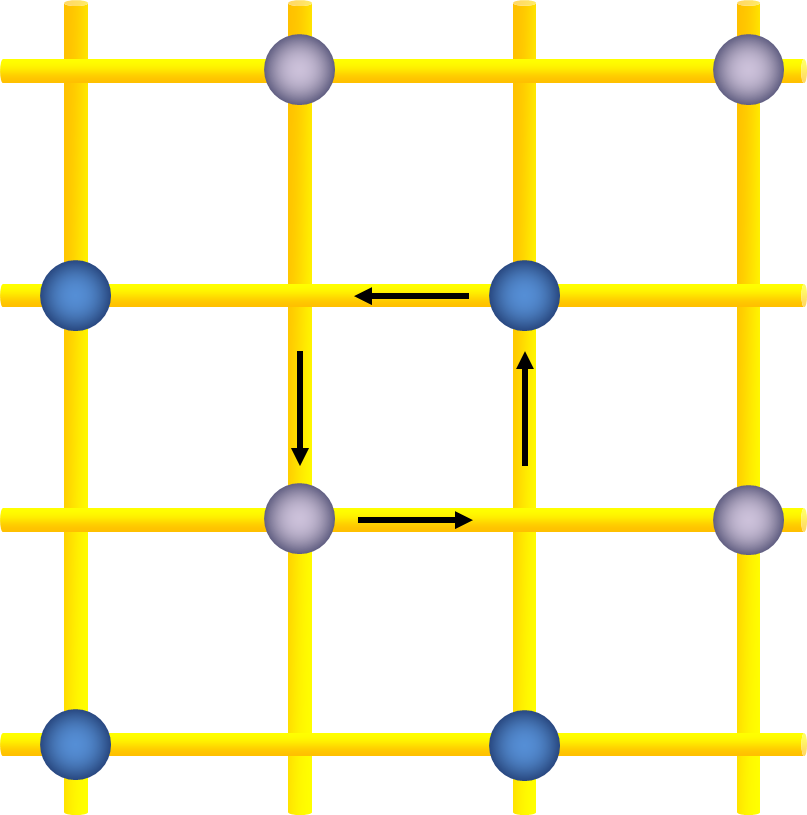}
\par\end{centering}

\caption{The loop method: Heavy, ancillary fermions (of two types, represented by the color) are constrained energetically to occupy certain vertices.
Virtual loops of such fermions (as in the middle plaquette), carrying the gauge degrees of freedom along with them, generate the plaquette terms.
For further details refer to \cite{AngMom}.}

\label{loopfig}
\end{figure}

In \cite{AngMom}, we have introduced the "loop method",
which uses heavy ancillary fermions - constrained to populate only
some special vertices that may move only
virtually, closing loops along plaquettes, carrying the gauge degrees
of freedom along the way (utilizing the on-link interactions (\ref{eq:gip}))
and thus effectively forming the plaquette interactions (see figure
\ref{loopfig}). Although these interactions are obtained as fourth-order
processes and might seem too weak, they are practically only second
order interactions, since the perturbative
series of the gauge-matter Hamiltonian contains only even orders
(fermions must hop virtually across an even distance, in order to finish
in a vertex in which they may stay)
and thus the fourth order is the second-leading one.

One can introduce yet other fermionic species (non-constrained) to
serve as "regular", dynamical fermions, on top of the ancillary
static ones.

We have utilized the loop method for a 2+1 dimensional compact $QED$
in the spin-gauge approach - plaquettes are constructed from the 1+1
dimensional link interactions, generated due to hyperfine angular
momentum conservation. This included a numerical proof of principle,
considering some gauge invariant but non-Kogut-Susskind corrections
to the Hamiltonian, due to the effective calculation in the truncated
case. It was shown that the effective Hamiltonian still gives rise
to the expected spectrum regardless of these corrections in several
regimes of the Hamiltonian parameter, also outside the strong coupling
limit.

The loop method has also been formulated and utilized for "ideal"
simulations, rather than the above described "realistic condition"
$cQED$ in 2+1 dimensions. This includes "ideal" (not truncated)
$cQED$, as well as $SU(N)$ gauge theory with unitary matrices (which
is not realistically achieved in the current simulation
method).

Finally, we shall comment on the simulation of the $\mathbb{Z}_N$ lattice gauge theory, whose relevance to $QCD$ confinement has
been stated above. Since this theory involves
finite Hilbert spaces on the links, one could, in principle, simulate
the exact model, with no truncations and approximations. An explicit
example for $\mathbb{Z}_{3}$ was given.
 Since the local Hilbert space is finite,
we use different bosonic levels on each link, playing the role of the different eigenstates
of the $\mathbb{Z}_N$ $P$ operator. The on-link interactions are with auxiliary fermions (which are not real $\mathbb{Z}_N$,
and are eliminated in the loop method), which can thus be also hard-core bosons (the statistics is irrelevant).
The gauge invariant operators are obtained using the conservation of hyperfine angular momentum in collisions, as before,
but the challenge here is to realize the cyclic "Escher staircase" nature of the group. This is done by hybridization of atomic
levels, as discussed in \cite{AngMom}.

Dynamical fermions may be included in these simulations as well, following the formulation of \cite{PRD}.

\subsection{Gauge invariance from atomic interaction symmetries - some general remarks}

To conclude, the approach of exact gauge invariance seems to be better
than the previous effective approach, as it is applicable to a larger class
of models (with the loop method for the plaquette interaction, for example).

This is since it does not have to include the Gauss's law constraints, which
are hard to realize (and even harder for non-Abelian gauge groups,
due to the non-commuting and complicated generators) and thus this
approach seems the preferred to proceed with.

Regarding the simulated theories: the simulation of $cQED$
in the non-effective approach seems closer to a physical
realization than the previous, effective ones.
A better physical realization of the non-Abelian models is required,
in order to
\begin{enumerate}
\item Allow for a better 1+1 simulation, with no need to use perturbation
theory at all (although the current realization does not rely on perturbation
theory for gauge invariance, only to the generation of links).
\item Include further nonperturbative regimes of $1+1$ dimensional theories.
\item Extend the simulations to 2+1 dimensions, in a way that plaquettes
will be generated using the loop method even out of gauge invariant
building blocks which contain non-unitary matrices; This should be possible using the truncation
scheme of \cite{PRD}, but as explained above, a concrete physical realization is still lacking.
\item Allow for simple generalizations to $SU\left(N\right)$ with $N>2$
and further gauge groups.
\end{enumerate}

\section{Discussion}

\begin{figure}[t]
\begin{centering}
\includegraphics[scale=0.7]{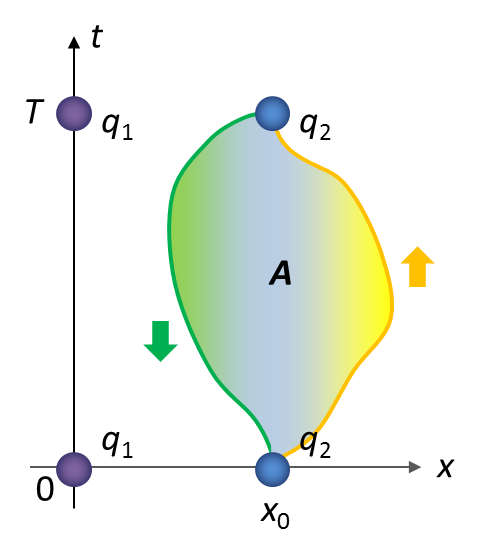}
\par\end{centering}

\caption{In the gedanken experiment presented in \cite{Topological}, whose realization was described in \cite{Zohar2013}. A meson, consisting of two
static quarks/charges, whose length is initially $x_0$, is brought into a superposition of two lengths, using two internal levels of one of them
(e.g. $\left|\uparrow\right\rangle$ and $\left|\downarrow\right\rangle$) which are affected by two different potentials, and thus
follow two different trajectories; At time $t$ the two mesons in superposition are brought to the same position again, resulting again with an $x_0$
sized meson. After performing a Ramsey spectroscopy, within a confining phase, the probabilities for being in either of the two external states
depend on the area enclosed by the trajectories $A$. This experiment may be used for detecting a confining phase, and within it - for measuring the
string tension. For a more detailed account of this procedure, the reader may refer to \cite{Topological}.}

\label{Wloop}
\end{figure}

The class of Abelian and non-Abelian lattice gauge theories that has been considered at the present work
constitutes the building blocks of the standard model of elementary particles, that
currently provides us with the best known description of high-energy physics phenomena.
The standard model has been so far extensively tested, and currently agrees with the experimental results from particle accelerators and cosmological observations,
 up to energy scales of $\sim 10^{12} \text{eV}$.
On the other hand, the physics which is used here to describe the many-body system of ultra-cold atoms is well understood and has been tested with high accuracy, down to energy scales of $\sim 10^{-7} \text{eV}$: $\sim 19$ orders of magnitude below the energy scales of the simulated physics!

Furthermore, the difference between the simulated system and analog-simulator is not only in the energy scales: there rather exists a striking difference in the formal structure and the physical principles that govern the physics of high-energy phenomena and ultracold atoms.
The atomic system at hand is non-relativistic, and seems to lack the sort of constrained symmetries and dynamical structure that are so essential for  relativistic HEP models. Unlike high energy physics, atomic dynamics are number conserving, and  manifest neither local gauge invariance, nor Lorentz invariance, which lead, among other things, to the familiar causal space-time structure, to charge conservation laws, to long-range interactions in $QED$ and to quark confinement in $QCD$.

So the problem at hand seems conceptually not obvious: can high-energy field theories be simulated by the physics of ultra low-energy systems?

It turns out, however and somewhat surprisingly, that there are several ways to manifest both Lorentz (as an effective long wavelength property) and local-gauge invariance. The latter can be obtained in two ways:
It may arise either as a property of a low energy sector of the atomic system - hence as an effective symmetry, or, alternatively, by re-arranging the interactions such that the given low-energy symmetries of the atomic collisions are converted into a form which is equivalent to local (Abelian or non-Abelian) gauge symmetries. The first case of "effectively emerging" gauge symmetry seems to provide yet another framework to study models in which local gauge invariance is not a fundamental property of the complete theory, but rather emerges as a low energy phenomenon as suggested in \cite{Nielsen1983}. The alternative "exact" mapping might, on the other hand, be more suitable for mimicking theories wherein local gauge invariance is a fundamental property.

In this report, we have reviewed  recent progress on quantum simulations of lattice gauge theories using
ultracold atoms, focusing on some of the works and approaches. These proposals serve as proofs of principle of the possibility to simulate high energy physics,
and even gauge theories, using ultracold atoms.

Our current results already suggest that simple enough models, such as compact-QED ($U(1)$), and possibly also more involved, non-Abelian $SU(2)$ models, manifesting exotic QCD effects, (such as quark confinement) could, indeed, be experimentally studied using current table-top cold atoms experimental methods.

\subsection{Advantages of quantum simulation over other approaches}

\begin{enumerate}
\item As fermions "come for free" in these atomic scenarios, one may
avoid the "sign problem" of Grassman integration and consider
regimes of finite chemical potential, such as the exotic color-superconductivity
and quark-gluon plasma phases of Quantum Chromodynamics \cite{McLerran1986,Kogut2004,Fukushima2011}.
\item Real time dynamics is possible here, rather than in the statistical
Monte-Carlo simulations (and this applies already for the current
proposals). This allows, or shall allow,

\begin{enumerate}
\item Observing real time dynamical phenomena, including the ones described
in the papers: deforming and breaking of flux tubes and loops, pair
creation etc.
\item Changing adiabatically the Hamiltonian parameters - finding ground
states of systems, by switching from {}``trivial'' regimes where
the ground state is known (e.g. no-interactions, strong limit) to
other regimes; Also probing for phase transitions and creating phase
diagrams of theories, depending on the theory's constants. This shall
be highly relevant for $QCD$ once feasible simulation systems are
proposed and realized.
\end{enumerate}
\item Possibility to consider finite temperature models as well\cite{Svetitsky,SvetitskyReview,Kogut2004}, including with real time dynamics .
\item Possibility to realize otherwise only gedanken experiments. For example, a realization of the gedanken experiment
proposed in \cite{Topological}, which
brings the idea of Ramsey spectroscopy \cite{Ramsey1990} into high
energy physics, suggesting a way to probe for the confining phase
and measure the string tension related with confinement (see figure \ref{Wloop}).
\end{enumerate}

\subsection{Simulated models}
As described in this report, a large class of Abelian and non-Abelian physical models can be simulated, including,
explicitly, the lattice gauge theories of
\begin{enumerate}
\item Compact $QED$ - Kogut Susskind and spin-gauge (truncated) models,
in 1+1 and 2+1 dimensions, with both effective and exact gauge invariance,
with or without dynamical matter. In 1+1 dimensions, the simulation
is possible with no perturbation theory at all.
\item $\mathbf{\mathbb{Z}}_{N}$ - so far, $\mathbf{\mathbb{Z}}_{3}$ has been extensively studied - 2+1 pure gauge, or with dynamical matter (following \cite{PRD}) - an exact quantum simulation of a non-truncated Hilbert space.
\item $SU(N)$ - so far $SU(2)$ in 1+1 dimensions, with limited applicability
to more dimensions has been studied in detail, while other $SU(N)$ groups still await extended study.
\end{enumerate}

The simulatable theories, as presented in this report, are summarized in the table below, where KS - Kogut-Susskind, trunc. - truncated, st.c . - Strong coupling, YM - Yang Mills.
\\

\begin{tabular}{|c|c|c|c|}
  \hline
      & 1+1 with matter & 2+1 Pure & 2+1 with matter \\
    \hline
  $U(1)$ & Full KS + trunc. & Full KS + trunc. & Full KS + trunc.  \\
  \hline
  $\mathbb{Z}_3$ & Full & Full & Full  \\
  \hline
  $SU(2)$ & YM + trunc. & YM + trunc. (st. c.) & YM + trunc. (st. c.) \\
  \hline
\end{tabular} \\

For other $\mathbb{Z}_N$ and $SU(N)$ groups, the simulation formalism still applies - however, experimental realizations seem highly challenging, and thus, perhaps, one should seek for other methods.

\subsection{Open problems}

These models and simulation proposals still require some more study, both from
theoretical and experimental aspects.

On the theoretical level, it is very important to understand the implications
of using truncated systems - are systems with a finite local Hilbert space
accurate enough for simulations of  Kogut-Susskind models, if one exploits
an appropriate coarse-graining scheme? And, resulting from that, what should be
the continuum limit of such theories, if it exists?

Such truncated models may be treated using tensor network techniques, and indeed,
some studies approaching lattice gauge theories with tensor networks have already
been carried out \cite{Banuls2013,Banuls2013a,Rico2013,Tagliacozzo2014,Kuhn2014,Haegeman2014,Silvi2014,Buyens2014,Saito2014,Banuls2015,Kuhn2015,Pichler2015,fPEPS2015}.

As far as the simulating systems are considered, the following should
be studied, in order to progress on the experimental side:
\begin{enumerate}
\item Techniques for generating complex lattice and superlattice structure
- both for more challenging simulation models and for increasing the
number of atoms which may populate a single site.
\item Improving the controllability of parameters - new ways to control
scattering coefficients, for example; Also, Strengthening the possible
interactions is important, for example for the realization of effective,
$\sim J^{2}/U$, interactions, or the second $SU(2)$ realization scheme
presented above, as well as, potentially, the use of the truncation scheme of \cite{PRD}
for the simulation of other non-Abelian theories (with either compact Lie or finite gauge groups),
including, of course $SU(3)$ for $QCD$.
\item Methods to create longer-living and more robust lattices - cooling
techniques, ways to deal with atomic losses and more.
\item Measurement techniques - the current single-addressing techniques
\cite{Bakr2009,Weitenberg2011,Parsons2015} require very sophisticated experimental
methods, limited to a small number of experimental groups. Development
of such, and other techniques, shall help boosting the quantum simulations
of complicated physical models.
\end{enumerate}

\subsection{Summary}
Atomic physics and high energy physics have, indeed, what to offer
to one another, suggesting a very interesting and exciting physical
research focusing on their interface. Reconstructing high energy physics
from atomic components offers a great lesson in understanding
the interactions unfolded within the standard model of particle physics.
And on the broader sense, the continuation of such works on quantum
simulations of high energy physics, and the growth of the community
working in this direction, may lead to useful new ways, methods and
results in understanding the most fundamental degrees of freedom in
our physical universe.

\subsection*{Acknowledgements}
E.Z. acknowledges the support of the Alexander-von-Humboldt Foundation. B.R. wishes to thank the hospitality of MPQ.
Part of the work has been supported by the EU Integrated Project SIQS.

\section*{References}
\bibliography{ref}

\end{document}